\documentclass[12pt,english,preprint]{revtex4-1}
\usepackage{times}
\usepackage{array}
\usepackage{longtable}
\usepackage{float}
\usepackage{amsmath}
\usepackage{bm}
\usepackage{graphicx}
\usepackage{amssymb}
\usepackage{color}
\usepackage{soul}
\usepackage[FIGTOPCAP]{subfigure}

\makeatletter
\usepackage{fancyhdr}
\usepackage{babel}
\makeatother
\begin{document}

\title{An Adjoint Formulation of Energetic Particle Confinement}

\author{Christopher J. McDevitt}
\email{cmcdevitt@ufl.edu}
\affiliation{Nuclear Engineering Program, Department of Materials Science and Engineering, University of Florida, Gainesville, FL 32611, United States of America}
\author{Jonathan S. Arnaud}
\affiliation{Nuclear Engineering Program, Department of Materials Science and Engineering, University of Florida, Gainesville, FL 32611, United States of America}

\date{\today}

\begin{abstract}

An adjoint formulation of energetic particle confinement in axisymmetric tokamak geometry is derived and evaluated using a physics-informed neural network (PINN). The PINN estimates the mean escape time of energetic ions by solving an inhomogeneous adjoint of the drift kinetic equation with a Lorentz collision operator, yielding predictions of fast ion loss in tokamak geometry due to direct ion orbit loss and collisional transport. To our knowledge, this is the first time a PINN has been used to solve the drift kinetic equation in tokamak geometry, a challenging problem due to the large time scale separation between the rapid transit time of energetic ions and their slow collisional time scale. It is shown that a careful and intentional design of a PINN is able to learn the mean escape time across the majority of the plasma volume, suggesting a path toward constructing a rapid surrogate for use within a broader optimization framework.

\end{abstract}

\maketitle


\section{Introduction}

Energetic particles are a ubiquitous aspect of magnetic fusion plasmas. Such particles emerge as a result of ion heating schemes including neutral beam injection~\cite{hemsworth2008status} and ion cyclotron resonance heating (ICRH)~\cite{adam1987review, kazakov2017efficient}. For a burning deuterium-tritium plasma, the confinement of energetic alpha particles is critical for the sustainment of the fusion plasma. In addition, when such particles escape they can cause substantial damage to plasma-facing components~\cite{fasoli2007physics, bonofiglo2024alpha}. Obtaining an understanding of the confinement time of such particles is thus vital to the operation of a magnetic fusion device.

In the present work, an initial step toward a framework through which the confinement time of an energetic ion or electron can be efficiently estimated is developed. While evaluating the confinement time of an energetic particle is often of greatest importance for non-axisymmetric magnetic fields arising from error fields~\cite{goldston1981confinement}, magnetic islands~\cite{carolipio2002simulations} energetic particle modes~\cite{heidbrink2020mechanisms}, or for a 3D stellarator equilibrium~\cite{galeev1969plasma}, this work will consider the simpler problem of an axisymmetric tokamak equilibrium. Our motivation for considering this simpler problem is to investigate the potential for evaluating metrics of energetic particle transport via the solution to an inhomogeneous adjoint problem. As described below, the inhomogeneous adjoint of the drift kinetic equation provides a direct means of computing the average confinement time of an ion or electron as a function of the particle's initial phase space position~\cite{sarkimaki2020efficient}, a quantity we will refer to as the mean escape time. While this adjoint problem can be solved by a variety of numerical methods, we will consider a physics-informed neural network (PINN)~\cite{raissi2019physics} in the present work. Physics-informed neural networks correspond to a machine learning approach capable of integrating both physics and data into the training of a neural network. A distinguishing feature of a PINN is that it can be extended to learn the parametric dependence of the solution of a PDE~\cite{sun2020surrogate}. Noting that optimizing the transport properties of energetic particle orbits often requires exploring a broad range of magnetic field configurations, obtaining a rapid surrogate capable of computing quantities of interest such as the mean escape time of energetic particles provides a powerful tool for assessing energetic particle transport.

Our aim in the present paper is not to provide such a parametric solution, but to instead investigate the ability of a PINN to solve the drift kinetic equation for an energetic ion population. The extension to a parametric solution will be treated in a future work. Specifically, while PINNs have previously been used to solve the relativistic~\cite{arnaud2024physics, mcdevitt2025primary, Arnaud2025arunaway} and non-relativistic~\cite{mcdevitt2024knudsen} Fokker-Planck equations, the large time scale separation between the rapid bounce or transit times and slow collision time of energetic particles in tokamak geometry, represents a far more challenging problem. Utilizing advanced optimization algorithms~\cite{urban2025unveiling, kiyani2025optimizer, wang2025gradient}, we aim to show that PINNs are able to robustly resolve several features of this challenging system, even in the absence of data. As described further below, as the energy of the ions is increased, leading to a greater time scale separation between the fast bounce/transit motion and slow collision time of energetic ions, our implementation of a PINN struggles to quantitatively describe the escape time of the best confined ion orbits, although it does resolve the solution in the outer region of the tokamak including phase space regions of prompt ion loss~\cite{hinton1985neoclassical}.


As an alternative means of evaluating the fast ion mean escape time, we introduce a new drift-kinetic solver based on the JAX framework. While in the present work this solver is only used to verify solutions from the PINN, future work will leverage particle data generated by this drift-kinetic solver to help accurately resolve the mean escape time of well confined ion orbits, along with facilitating the training of the PINN over a broad range of plasma parameters. This extended PINN framework, trained using both physics and particle data, would thus provide an efficient surrogate for fast ion confinement appropriate for use inside optimization loops.

The remainder of this paper is organized as follows. Section \ref{sec:ADKE} derives an adjoint problem for the mean escape time of an energetic particle. Brief overviews of the physics-informed deep learning framework and JAX particle-based solver are given in Sec. \ref{sec:PCDL}. The mean escape time of an energetic ion species is computed in Sec. \ref{sec:ET}. Conclusions and a discussion of future directions are given in Sec. \ref{sec:C}

\section{\label{sec:ADKE}Adjoint of Steady State Drift Kinetic Equation}

In this section we will define an adjoint problem based on the inhomogeneous steady state drift kinetic equation. Our starting point will be to define the Green's function $F_s \left(\mathbf{z}; \mathbf{z}_0 \right)$ for the steady state drift kinetic equation~\cite{Helander-Sigmar:book}:
\begin{subequations}
\label{eq:ADKE1}
\begin{equation}
\nabla \cdot \left( \mathbf{\dot{X}} F_s\right) + \frac{\partial}{\partial v_\Vert} \left( \dot{V}_\Vert F_s \right) -  C_s \left( F_s \right) = \delta \left( \mathbf{z} - \mathbf{z}_0 \right)
, \label{eq:ADKE1a}
\end{equation}
where $\mathbf{z} = \left( \mathbf{x}, v_\Vert, \mu \right)$ is the phase space position of the particle and the characteristic equations are given by:
\begin{equation}
\mathbf{\dot{X}} = v_\Vert \bm{\hat{b}} + \mathbf{v}_{EB} + \mathbf{v}_{Ds}
, \label{eq:ADKE1b}
\end{equation}
\begin{equation}
\dot{V}_\Vert = \frac{q_s}{m_s} \bm{\hat{b}} \cdot \mathbf{E} - \mu \bm{\hat{b}} \cdot \nabla B
. \label{eq:ADKE1c}
\end{equation}
\end{subequations}
Here, the subscript $s$ is the species index, $\mu \equiv v^2_\perp/\left( 2B \right)$ is the magnetic moment, $v_\Vert$ is the velocity parallel to the magnetic field, $C_s \left( F_s \right)$ is a collision operator, taken to be a Lorentz collision operator
\begin{equation}
C_s \left( F_s \right) = \frac{\nu_D}{2} \frac{\partial}{\partial \xi} \left[ \left( 1-\xi^2 \right) \frac{\partial F}{\partial \xi} \right]
, \label{eq:ADKE1sub2}
\end{equation}
and we have defined the $\mathbf{E} \times \mathbf{B}$ and magnetic drifts as
\begin{subequations}
\begin{equation}
\mathbf{v}_{EB} = \frac{\mathbf{E} \times \mathbf{B}}{B^2}
, \label{eq:ADKE2a}
\end{equation}
\begin{equation}
\mathbf{v}_{Ds}= \frac{\left( v^2_\Vert + \mu B\right)}{\omega_{cs}} \bm{\hat{b}} \times \nabla \ln B
. \label{eq:ADKE2b}
\end{equation}
\end{subequations}
The magnetic drifts have been simplified by the approximation $\nabla \times \mathbf{\hat{b}} \approx \mathbf{\hat{b}} \times \nabla \ln B$, appropriate in the limit of a ``vacuum plasma.'' As boundary conditions we will enforce $F_s = 0$ at the spatial boundaries whenever $U_r \equiv \mathbf{r} \cdot \mathbf{\dot{X}} < 0$, where $\mathbf{\hat{r}}$ is the radial unit vector, and we have assumed a circular spatial boundary, but leave $F_s$ unconstrained for $U_r \equiv \mathbf{r} \cdot \mathbf{\dot{X}} > 0$, i.e.
\begin{equation}
F_s \left( r=r_{wall}\right) = 
\begin{cases}
0, & U_r < 0 \\
\text{unconstrained}, & U_r \geq 0 
\end{cases}
, \label{eq:ADKE2sub0}
\end{equation}
where $r_{wall}$ is the location of the surrounding wall, taken as the plasma minor radius $a$, for simplicity. With these boundary conditions particles are not allowed to enter the spatial domain, but particles are able to exit. The resulting Green's function can thus be physically interpreted as the steady state particle distribution resulting from a unit source located at $\mathbf{z} = \mathbf{z}_0$.

To derive an adjoint problem for the mean escape time of an energetic particle, we begin by multiplying the left hand side of Eq. (\ref{eq:ADKE1a}) by the solution to the adjoint equation $T_s$, which after 
successive integrations by parts, yields:
\begin{align}
\int d^3v d^3 x T_s & \left[ \nabla \cdot \left( \mathbf{\dot{X}}F_s \right) + \frac{\partial}{\partial v_\Vert} \left( \dot{V}_\Vert F_s \right) - C_s \left( F_s \right) \right] \nonumber \\ 
& = \int d^3v d^3 x F_s \left[ -\mathbf{\dot{X}} \cdot \nabla T_s - \dot{V}_\Vert \frac{\partial T_s}{\partial v_\Vert} - C^*_s \left( T_s \right) \right] \nonumber \\ 
& + \int d^3v \int d \Gamma \left. \left[ U_r T_s F_s \right] \right|_{r=r_{wall}} 
. \label{eq:ADKE2sub1}
\end{align}
Here, the first term on the right hand side (RHS) is the Green's function times the adjoint of the steady state drift kinetic equation, and the second term on the RHS describes fluxes through the spatial boundary $\Gamma$. The boundary condition on the Green's function $F_s$ [Eq. (\ref{eq:ADKE2sub0})] implies the spatial boundary term will vanish when $U_r < 0$, where setting the boundary conditions on $T_s$ when $U_r \geq 0$, allows for distinct adjoint problems to be defined. 

An adjoint problem for the mean escape time~\cite{grasman2013asymptotic, Liu:2017} of an energetic particle can be derived by considering the inhomogeneous adjoint equation
\begin{equation}
\mathbf{\dot{X}} \cdot \nabla T_s + \dot{V}_\Vert \frac{\partial T_s}{\partial v_\Vert} + C^*_s \left( T_s \right) = -1
. \label{eq:IADK1}
\end{equation}
Substituting Eq. (\ref{eq:IADK1}) along with the steady state Green's function defined by Eq. (\ref{eq:ADKE1a})
into the right and left hand sides, respectively, of Eq. (\ref{eq:ADKE2sub1}), yields
\begin{align}
T_s \left( \mathbf{x}_0, \mathbf{v}_0\right) &= N_s + \int d \Gamma \int d^3v \left. \left[ U_r T_s F_s \right] \right|_{r_{wall}}
, \label{eq:IADK2}
\end{align}
where $\Gamma$ is the surface of the surrounding wall and $N_s\equiv \int d^3v \int d^3x F_s$ is the number of particles remaining in the domain due to the unit source on the right hand side of Eq. (\ref{eq:IADK1}). At the spatial boundary we will take
\begin{equation}
T_s \left( r=r_{wall}\right) = 
\begin{cases}
\text{unconstrained}, & U_r \leq 0 \\
0, & U_r > 0 
\end{cases}
. \label{eq:ADKE2suba}
\end{equation}
This boundary condition follows since a particle born on the spatial boundary with an outward directed velocity would be lost immediately, such that $T_s=0$, however, for an inward directed velocity this particle would not be immediately lost. For the boundary condition defined by Eq. (\ref{eq:ADKE2suba}), and noting that $F_s = 0$ for $U_r<0$ [Eq. (\ref{eq:ADKE2sub0})], yields
\begin{align}
T \left( \mathbf{x}_0, \mathbf{v}_0\right) &= N_s
. \label{eq:IADK3}
\end{align}
Further noting that $N_s$ will scale directly with the confinement time of the particles, $T_s\left( \mathbf{x}_0, \mathbf{v}_0\right)$ will act as a measure of how well particles injected at a given phase space location are confined on average, a quantity often referred to as the mean escape time or expected exit time~\cite{grasman2013asymptotic}. We note that for the present example, all ions will escape for $t\to \infty$, such that the mean escape time will be finite throughout the domain.

\section{\label{sec:PCDL}Physics-Constrained Deep Learning Framework}

The primary focus of the remainder of this work will be the solution of the adjoint problem derived in Sec. \ref{sec:ADKE}. A brief overview of our implementation of a PINN is given in Sec. \ref{sec:PINN}, where the reader is referred to Refs. \cite{cuomo2022scientific, wang2023expert, toscano2025pinns} for additional information on this rapidly expanding area. In addition, a newly developed GPU accelerated particle-based solver for the drift kinetic equation is described in Sec. \ref{sec:PBDK}. For the present work, this particle-based solver will be used to validate the PINN's solutions to the drift kinetic equation.

\subsection{\label{sec:PINN}Physics-Informed Neural Networks}

Physics-constrained machine learning methods~\cite{karniadakis2021physics} seek to incorporate physical constraints into the training of a neural network. Physics-informed neural networks correspond to a prominent example of this area. An attractive feature of PINNs is that they can be used to solve PDEs and ODEs directly in the absence of data~\cite{van1995neural, raissi2019physics}. They also provide a mechanism through which physical constraints can be used to enrich sparse data sets~\cite{cai2021physics, mathews2022deep, mcdevittSciTech2024}. In its simplest form, the loss function of a PINN can be written as:
\begin{equation}
\text{Loss} = \frac{1}{N_{data}} \sum^{N_{data}}_{i=1} \left[ T_i - T \left( \mathbf{z}_i; \bm{\lambda}_i \right)\right]^2 + \frac{w_{pde}}{N_{PDE}} \sum^{N_{PDE}}_{i=1} \mathcal{R}^2 \left( \mathbf{z}_i; \bm{\lambda}_i \right)
, \label{eq:PINN1}
\end{equation}
where $T$ is the quantity we seek to predict, the mean escape time for the current work, $\mathcal{R}$ is the residual of the governing PDE, $w_{PDE}$ is a scalar weight applied to the PDE bias of the loss, $\mathbf{z} = \left( \mathbf{x}, \mathbf{v} \right)$ are the independent variables, and $\bm{\lambda}$ describes physical parameters, such as collisionality or inverse aspect ratio. The first term represents a bias due to data, where the data may be boundary conditions, or synthetic or experimental data. The second term is a bias due to the governing equations, which for the present work will be the inhomogeneous adjoint of the steady state drift kinetic equation. In the absence of synthetic or experimental data, the first term will only contain the boundary conditions. Successfully minimizing the loss will thus correspond to identifying a solution to the PDE while satisfying the boundary conditions. Unlike traditional PDE solvers, a parametric solution to a PDE can be straightforwardly obtained by training across a broad range of physics parameters $\bm{\lambda}$. Thus, while the training time of such a model may be substantially longer than a traditional PDE solver, once trained, the fast inference time of a PINN allows for fast online prediction times across a broad range of plasma conditions. This property can be used for the development of fast surrogate models, where some recent examples include runaway electron generation~\cite{McDevitt:hottail:2023, Arnaud2025arunaway}, MHD equilibrium~\cite{jang2024grad}, plasma thrusters~\cite{luo2025parametric}, or non-thermal ion distributions~\cite{mcdevitt2024knudsen}. 

Key elements in obtaining accurate results from PINNs correspond to (i) the identification of an appropriate neural network architecture, (ii) the phase space weighting of the residual, and (iii) the selection of an efficient optimization algorithm for minimizing the loss. Regarding item (i) we will employ a fully connected feedforward neural network, but with an `output layer' that enforces positivity as a hard constraint. In so doing, this both prevents certain unphysical solutions from being predicted by the PINN, and perhaps more importantly, narrows the space of solutions that the optimizer searches across. Defining $T_{NN}$ to be the output of the hidden layers of the neural network, a simple transformation that enforces positivity is given by
\begin{equation}
T_s = \exp \left( T_{NN}\right)
. \label{eq:PF1}
\end{equation}
In addition to enforcing positivity, the exponential dependence of $T_s$ on the output of the hidden layers $T_{NN}$ provides a natural means of capturing the large disparity in escape times for energetic ions. Specifically, ions born near the plasma core are expected to be well confined, where they will only be lost by the relatively slow process of collisional transport, in contrast to those born near the edge, which may be lost on a very short time scale by direct orbit loss.

An additional challenge that arises is ensuring that the PINN is able to robustly connect the boundary condition Eq. (\ref{eq:ADKE2suba}) applied at the edge of the tokamak, to the solution throughout the core plasma. Noting that we expect the escape time $T_s$ to be small near the edge, training points in this edge region can be given greater emphasis by weighting the residual of the PDE by:
\begin{equation}
\mathcal{F} \left( \mathbf{z}; \bm{\lambda} \right) = \frac{A}{A+T_s\left( \mathbf{z}; \bm{\lambda} \right) }
, \label{eq:PF2}
\end{equation}
where $A$ is a constant, such that the loss can be written as:
\begin{equation}
\text{Loss} = \frac{1}{N_{bdy}} \sum^{N_{bdy}}_{i=1} \left[ T_{s,i} - T_s \left( \mathbf{z}_i; \bm{\lambda}_i \right) \right]^2 + \frac{w_{pde}}{N_{PDE}} \sum^{N_{PDE}}_{i=1}\left[ \mathcal{F} \left( \mathbf{z}_i; \bm{\lambda}_i \right) \mathcal{R} \left( \mathbf{z}_i; \bm{\lambda}_i \right) \right]^2
. \label{eq:PF3}
\end{equation}
Here, the boundary conditions defined by Eq. (\ref{eq:ADKE2suba}) are enforced by the first term, $\mathcal{R}$ is the residual of the inhomogeneous adjoint defined by Eq. (\ref{eq:IADK1}), and the choice of $A$ sets how much emphasis is given to the edge region. In particular, noting that near the edge $T_s$ will be of the order of a transit time, since ions on loss orbits will directly escape from the plasma, if $A$ is chosen to satisfy $A \ll1$, Eq. (\ref{eq:PF2}) will scale as $\propto 1/T_s$, such that the edge region will be heavily weighted, but the inner region of the plasma with $T_s \gg 1$, will receive minimal weighting. We thus expect this choice to allow for the outer region of the plasma to be accurately described, at the expense of poor accuracy in the inner region. In contrast, by choosing a larger value of $A$, improved accuracy for the inner region can be achieved, at the expense of having a looser connection to the edge boundary condition, which is vital for achieving a physically meaningful solution. We have found that values of $A$ between 10 and 100 allow for robust convergence of the PINN, where all results in the present paper are for $A=100$. In addition, the relative weight of the PDE residual compared to the boundary condition is set by the scalar $w_{PDE}$ weighing the first term in Eq. (\ref{eq:PF3}). We have found $w_{PDE}=100$ provides relatively rapid convergence of the loss.

The last key component of our PINN implementation will be the choice of an effective optimization algorithm. While the ADAM optimizer~\cite{kingma2014adam} is used ubiquitously across a range of machine learning applications, this first order optimization algorithm often is not able to achieve low losses for challenging PDEs. A common strategy is thus to use ADAM for the first phase of training, after which limited memory BFGS (L-BFGS)~\cite{liu1989limited}, a second order optimization algorithm, is used to more tightly converge the loss. Motivated by Refs. \cite{wang2025gradient} and \cite{urban2025unveiling, kiyani2025optimizer}, we have adopted a different optimization strategy that we have found to yield substantially improved accuracy. The first optimizer that we apply corresponds to the quasi-second order optimizer SOAP~\cite{vyas2024soap}. This scalable optimizer exhibits similar computational performance as ADAM, but can yield losses up to an order of magnitude smaller than those achievable by ADAM across a broad range of challenging applications of PINNs~\cite{wang2025gradient}. For the second stage of optimization, we utilize the Self-Scaled Broyden (SSBroyden) method~\cite{al2014broyden}. Like L-BFGS, SSBroyden approximates the Hessian, but introduces additional degrees of freedom into the optimization algorithm that have yielded substantial improvements in training across a range of complex PDEs~\cite{urban2025unveiling, kiyani2025optimizer}. While each training epoch is substantially more computationally demanding compared to ADAM or SOAP for the SciPy implementation we are currently employing, we have found it results in substantially faster convergence of the loss for the adjoint problem treated in this work. The primary limitations of SSBroyden as an optimization routine is that the SciPy implementation used in the present work does not efficiently utilize GPUs or support mini-batching. These limitations substantially increase the training time of the SSBroyden phase of training, and limit the number of training points employed.

\subsection{\label{sec:PBDK}JONTA: A Particle Based Drift Kinetic Solver}

Particle-based drift kinetic simulations of fast ion orbits are computationally intensive~\cite{white1984hamiltonian, hirvijoki2014ascot} requiring substantial computational resources and efficient software. While this has predominately been done with Fortran or C++ libraries, with GPU acceleration facilitated by libraries such as Kokkos~\cite{trott2021kokkos}, maintaining the support of these solvers can be cumbersome, due to its dependency on external backend libraries to effectively utilize high performance computing (HPC) hardware. Perhaps most importantly, given that PINNs are trained on GPUs and libraries with a large user base, this motivates the development of a traditional drift kinetic solver that runs on GPUs and utilizes the same libraries that are used by PINNs, facilitating the operation of the broader framework that contains an interplay between the PINN and a drift kinetic solver.

With the aforementioned discussion, we present Just anOther fuNcTionAl pusher (JONTA), which is built on the PyTorch~\cite{paszke2019pytorch} and JAX~\cite{bradbury2018jax} libraries that run on GPU accelerated hardware. For the present work, we will utilize the JAX backend, and employ a four stage Runge Kutta (RK4) integration scheme, which is part of a broader suite containing differentiable integrator methods in the Optax library~\cite{optax2020}. Guiding-center equations are evolved with RK4, and collisions are implemented with a Monte-Carlo operator described below. The version of JONTA used in this manuscript, together with the PINN training script are available on github.com/cmcdevitt2/DeepPlasma.


JONTA will be used to solve the guiding center equations defined by Eq. (\ref{eq:ADKE1}) together with a Monte Carlo collision operator describing pitch-angle scattering. The guiding center equations can be written as
\begin{subequations}
\label{eq:PBDK0}
\begin{equation}
\dot{X} = \hat{\mathbf{x}} \cdot \dot{\mathbf{X}}
, \label{eq:PBDK0a}
\end{equation}
\begin{equation}
\dot{Z} = \hat{\mathbf{z}} \cdot \dot{\mathbf{X}} 
, \label{eq:PBDK0b}
\end{equation}
\begin{equation}
\dot{v} = \frac{\partial v}{\partial v_\Vert} \frac{dv_\Vert}{dt} + \frac{\partial v}{\partial \mu} \frac{d\mu}{dt} + \frac{\partial v}{\partial \mathbf{X}} \cdot \frac{d\mathbf{X}}{dt}
, \label{eq:PBDK0c}
\end{equation}
\begin{equation}
\dot{\xi} = \frac{\partial \xi}{\partial v_\Vert} \frac{dv_\Vert}{dt} + \frac{\partial \xi}{\partial \mu} \frac{d\mu}{dt} + \frac{\partial \xi}{\partial \mathbf{X}} \cdot \frac{d\mathbf{X}}{dt}
, \label{eq:PBDK0d}
\end{equation}
\end{subequations}
where we have chosen cylindrical coordinates $\left( R, Z \right)$, defined $X \equiv R-R_0$, and made the transformation from $\left( \mathbf{x}, v_\Vert, \mu \right)$ to $\left( \mathbf{x}, \xi, v \right)$. Assuming an axisymmetric tokamak geometry with circular flux surfaces and noting the relations $\xi^2 =v^2_\Vert / \left( v^2_\Vert + 2\mu B\right)$ and $v^2=v^2_\Vert + 2\mu B$, Eq. (\ref{eq:PBDK0}) can be written as 
\begin{subequations}
\label{eq:PBDK1}
\begin{equation}
\dot{X} = \hat{\mathbf{x}} \cdot v_\Vert \bm{\hat{b}} = - \xi v \frac{Z}{qR_0}
, \label{eq:PBDK1a}
\end{equation}
\begin{equation}
\dot{Z} = \hat{\mathbf{z}} \cdot v_\Vert \bm{\hat{b}} + \hat{\mathbf{z}} \cdot \mathbf{v}_{Ds} = \xi v \frac{X}{qR_0} - \frac{1}{2} \rho^*_i \left( 1 + \xi^2 \right) v^2 \frac{a}{R_0}
, \label{eq:PBDK1b}
\end{equation}
\begin{equation}
\dot{v} = 0
, \label{eq:PBDK1c}
\end{equation}
\begin{equation}
\dot{\xi} = \frac{\partial \xi}{\partial v_\Vert} \frac{dv_\Vert}{dt} + \frac{\partial \xi}{\partial \mathbf{X}} \cdot \frac{d\mathbf{X}}{dt} = -\frac{1}{2} \left( 1 - \xi^2 \right) v \frac{Z}{qR_0} \frac{a}{R_0} \frac{B}{B_0}
. \label{eq:PBDK1d}
\end{equation}
\end{subequations}
Here, we normalized time to $a/v_{Ti}$, where $a$ is the plasma minor radius taken to be equal to $r_{wall}$, speed $v$ has been normalized to $v_{Ti}$, space to $a$, and we have defined $\rho^*_i = \rho_i/a$, $\rho_i = v_{Ti}/\omega^{(0)}_{c}$, $\omega^{(0)}_c = ZeB_0/m_i$, $B_0$ is the magnetic field on axis, and $q$ is the safety factor. For convenience we have neglected the electric field, and assumed the limit of a small inverse aspect ratio and low-$\beta$ plasma. Specifically, the magnetic field was taken to have the form:
\begin{equation}
\mathbf{B} = I \left( \psi \right) \nabla \varphi + \nabla \varphi \times \nabla \psi
, \label{eq:ADKE7}
\end{equation}
\begin{equation}
B_\varphi \left( r, \theta\right) = \frac{B_0}{1+ \epsilon \cos \theta}
, \label{eq:ADKE8}
\end{equation}
\begin{equation}
B_\theta \left( r, \theta\right) = \frac{B_\theta \left( r\right)}{1+ \epsilon \cos \theta}
, \label{eq:ADKE9}
\end{equation}
with
\begin{equation}
R = R_0 + r \cos \theta
, \label{eq:ADKE10}
\end{equation}
\begin{equation}
q = \frac{r B_0}{R_0 B_\theta \left( r \right)}
, \label{eq:ADKE11}
\end{equation}
and the poloidal flux function is related to the poloidal magnetic field by $d\psi /d r = R_0 B_\theta \left( r \right)$. The magnitude of the total magnetic field can be conveniently written as
\[
\frac{B \left( r,\theta \right)}{B_0} = \frac{R_0}{R} \sqrt{1+ \left( \frac{r}{q R_0} \right)^2} \approx \frac{R_0}{R}
,
\]
where in the last line we take $\left( a/R_0 \right)^2 \ll 1$. 

Equations (\ref{eq:PBDK1a}) and (\ref{eq:PBDK1b}) describe the spatial evolution of the energetic ions due to parallel streaming and a vertical drift arising from the grad-B and curvature drifts.
Our motivation for neglecting the electric field is that in this initial study we will assume stationary and homogeneous density, temperature and current profiles, such that no electric field is expected to arise. In the absence of the electric field, the collisionless velocity space dynamics are set by the mirror force, which only impacts the pitch of the ions such that the speed $v$ will be constant. While a more complete set of guiding center equations can be easily adopted, for this initial study, Eq. (\ref{eq:PBDK1}) will prove sufficient to capture several non-trivial aspects of the collisionless orbit of energetic ions. 

\begin{figure}
\begin{centering}
\subfigure[]{\includegraphics[scale=0.5]{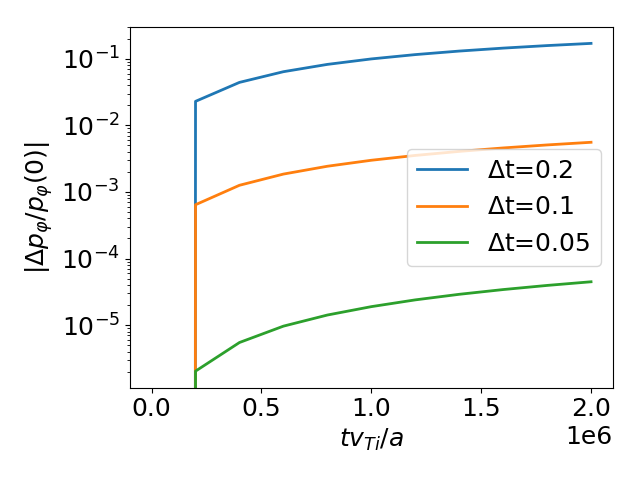}}
\subfigure[]{\includegraphics[scale=0.5]{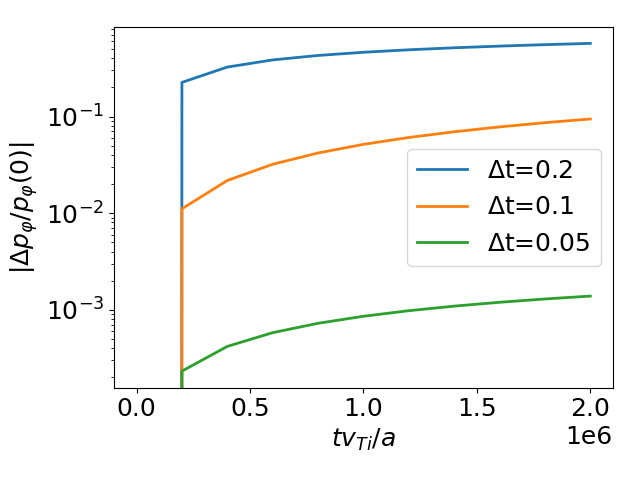}}
\par\end{centering}
\caption{Change of toroidal canonical momentum for 20 keV ions (panel a.) and 50 keV ions (panel b). Ten million deuterium ions were initialized randomly across the spatial and pitch domains. The time step was varied from $\Delta t = 0.2$ (blue curve), to $\Delta t = 0.1$ (orange curve), and $\Delta t = 0.05$ (green curve). The tokamak was assumed to have a minor radius of $a=0.5\;[\text{m}]$, an inverse aspect ratio of $a/R_0 = 1/3$, a magnetic field of $B_0 = 2\;[\text{T}]$, and a constant safety factor $q=2$.}
\label{fig:PBDK1}
\end{figure}

Aside from the speed $v$, this reduced set of guiding center equations conserves toroidal canonical momentum in the absence of collisions which can be written as:
\begin{equation}
p_\varphi = m_i R \frac{B_\varphi}{B} v_\Vert - Ze \psi
, \label{eq:PBDK2}
\end{equation}
For circular flux surfaces, and making analogous approximations as the guiding center equations [Eq. (\ref{eq:PBDK1})], allows the normalized toroidal canonical momentum to be expressed as:
\begin{equation}
\frac{p_\varphi}{R_0m_iv_{Ti}} = \frac{R}{R_0} \xi v - \frac{1}{2}\frac{1}{\rho^*_i} \frac{a}{R_0} \frac{r^2}{q}
, \label{eq:PBDK3}
\end{equation}
where we have taken the safety factor to be a constant for simplicity. Figure \ref{fig:PBDK1} provides a numerical demonstration of the conservation of toroidal canonical momentum conservation for ions with 20 and 50 keV. Here, while the error in toroidal canonical momentum conservation grows with time, after a time $2\times 10^6 a/v_{Ti}$, the error remains below $10^{-2}$ for a time step of $\Delta t = 0.1 a/v_{Ti}$, for 20 keV ions, and $\Delta t = 0.05 a/v_{Ti}$, for 50 keV ions. An error of less than one percent will be sufficiently accurate for the present study. JONTA simulations described in the remainder of this work will use a time step of $\Delta t = 0.1 a/v_{Ti}$ when studying 20 keV ions, and $\Delta t = 0.05 a/v_{Ti}$ for 50 keV ions.

The collision operator is taken to be a Lorentz operator [Eq. (\ref{eq:ADKE1sub2})], where the deflection frequency in the limit of $v \gg v_{Ti}$ can be expressed as:
\begin{equation}
\frac{a\nu_D}{v_{Ti}} = \left( a\frac{\hat{\nu}_{si}}{v_{Ti}} \right) \frac{v^3_{Ti}}{v^3} = \pi \left( n_i r^2_e a\right) Z^2_s Z^2_i \left( \frac{m_e c^2}{T_i} \right)^2 \ln \Lambda \frac{v^3_{Ti}}{v^3}
, \label{eq:PBDK4}
\end{equation}
where $r_e = 2.8179\times 10^{-15}\;[\text{m}]$ is the classical electron radius, $Z_i$ is the charge state of the background ions, and $Z_s$ is the charge state of the energetic ion. The Lorentz collision operator is implemented by the Monte Carlo equivalent~\cite{Boozer:1981}
\begin{equation}
\xi_{n+1} = \xi_n \left( 1 - \nu_D \Delta t \right) \pm \sqrt{\left( 1 - \xi^2_n \right) \nu_D \Delta t}
, \label{eq:PBDK5}
\end{equation}
where $\Delta t$ is the collisional time step, which is taken to be the same as the collisionless time step. While the Lorentz collision operator does not account for the slowing down of energetic ions, a critical aspect of describing energetic particle evolution, pitch-angle scattering will result in collisional cross field transport. 
The addition of a more complete set of physical processes including 3D magnetic fields, energetic particle modes, and collisional slowing down will be pursued in future work.

\begin{figure}
\begin{centering}
\subfigure[]{\includegraphics[scale=0.33]{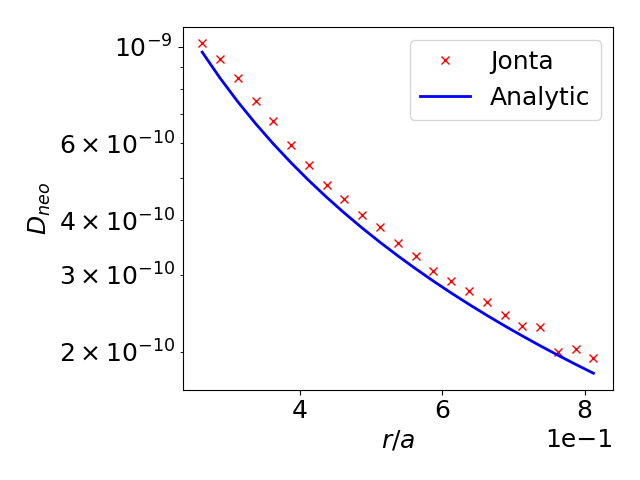}}
\subfigure[]{\includegraphics[scale=0.33]{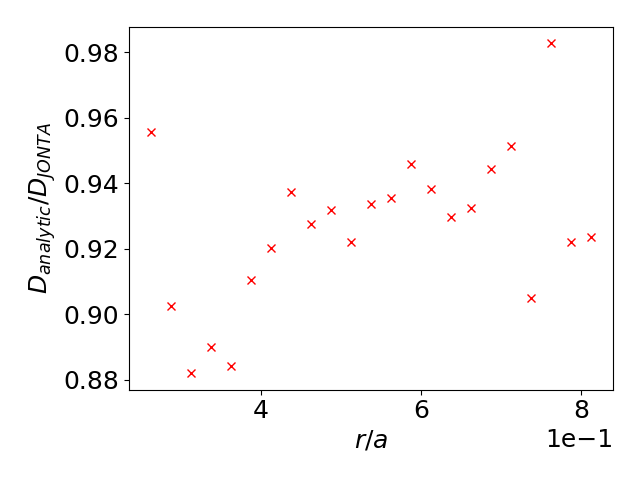}}
\subfigure[]{\includegraphics[scale=0.33]{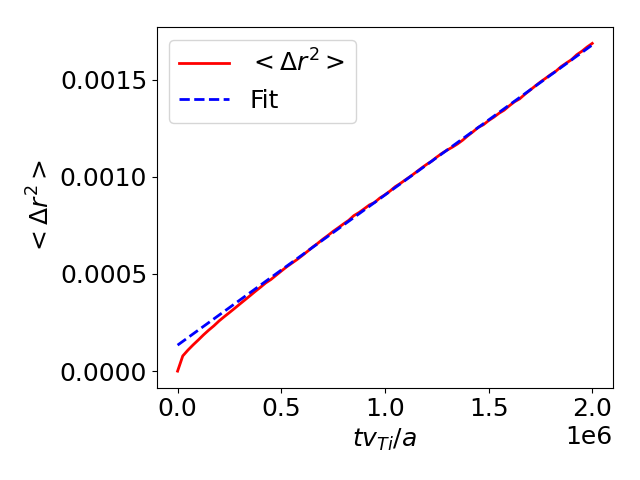}}
\par\end{centering}
\caption{(a) Comparison of neoclassical diffusivity computed from JONTA (red `x' markers) and the analytic expression given by Eq. (\ref{eq:PBDK6}) (solid blue curve). (b) Ratio of the analytic expression for the neoclassical diffusivity to JONTA's prediction. (c) Example fit of squared radial ion displacement versus time (dashed blue curve) with the numerically computed evolution (solid red curve) for particles with an initial radius of $\approx 0.51$. One million total marker particles were used, randomly distributed over pitch and radius. Spatially resolved estimates of transport were made by binning marker particles based on their initial location. The parameters were chosen to be $a=2\;[\text{m}]$, $a/R_0 = 1/6$, $B_0 = 10\;[\text{T}]$, $n_i = 10^{18}\;[\text{1/m}^3]$, $q=2$, $\ln \Lambda = 15$, and the ions were assumed to have an energy of $20\;[\text{keV}]$.}
\label{fig:PBDK2}
\end{figure}

To verify our implementation of the Monte Carlo collision operator we compared JONTA predictions of neoclassical transport with an analytic expression for the neoclassical diffusivity. Using the radial coordinate $r=\sqrt{X^2+Z^2}$, rather than the toroidal flux $\psi_t=r^2 B_0/2$, the expression for the neoclassical diffusivity derived in Refs. \cite{Rosenbluth:1972, Boozer:1981} can be written as:
\begin{equation}
\frac{D_{neo}}{av_{Ti}} = 0.689 \sqrt{2\varepsilon} q^2 \left( \frac{R_0}{r} \right)^2 \left( \frac{m_s v^2}{2T_i} \right) \left( \frac{a \nu_D}{v_{Ti}} \right)
, \label{eq:PBDK6}
\end{equation}
where $\varepsilon = r/R_0$, and $\nu_D$ is the pitch-angle scattering frequency defined in Eq. (\ref{eq:PBDK4}). To estimate the neoclassical transport predicted by JONTA we will consider a large tokamak device at low density. Specifically, we will take the density to be $n_i=10^{18}\;[1\text{/m}^3]$, the minor radius $a=2\;[\text{m}]$, a small inverse aspect ratio of $a/R_0=1/6$, and  a large toroidal magnetic field of $B_0 = 10\;[\text{T}]$. By considering a low density plasma this ensures the system is in the banana transport regime throughout the majority of the device, whereas the large device size and strong magnetic field results in the banana orbits of the fast ions being much smaller than the system size, a necessary condition for estimating an average diffusivity. A comparison of the neoclassical diffusivity estimated by JONTA with Eq. (\ref{eq:PBDK6}) is shown as a function of minor radius in Fig. \ref{fig:PBDK2}. Here, good agreement is evident between the computed neoclassical diffusivity and the approximate analytic expression. In particular, the maximum deviation is roughly 12\%, with a typical deviation less than 10\%. Note that we have removed the inner 20\% of the plasma radius from the comparison, since the de-trapping time scales with $\nu_D/\varepsilon$, and is thus very short near the magnetic origin, such that this region is not in the banana transport regime.

\subsection{\label{sec:EIC}Example Ion Orbits and Coordinate Conventions}

\begin{figure}
\begin{centering}
\subfigure[]{\includegraphics[scale=0.5]{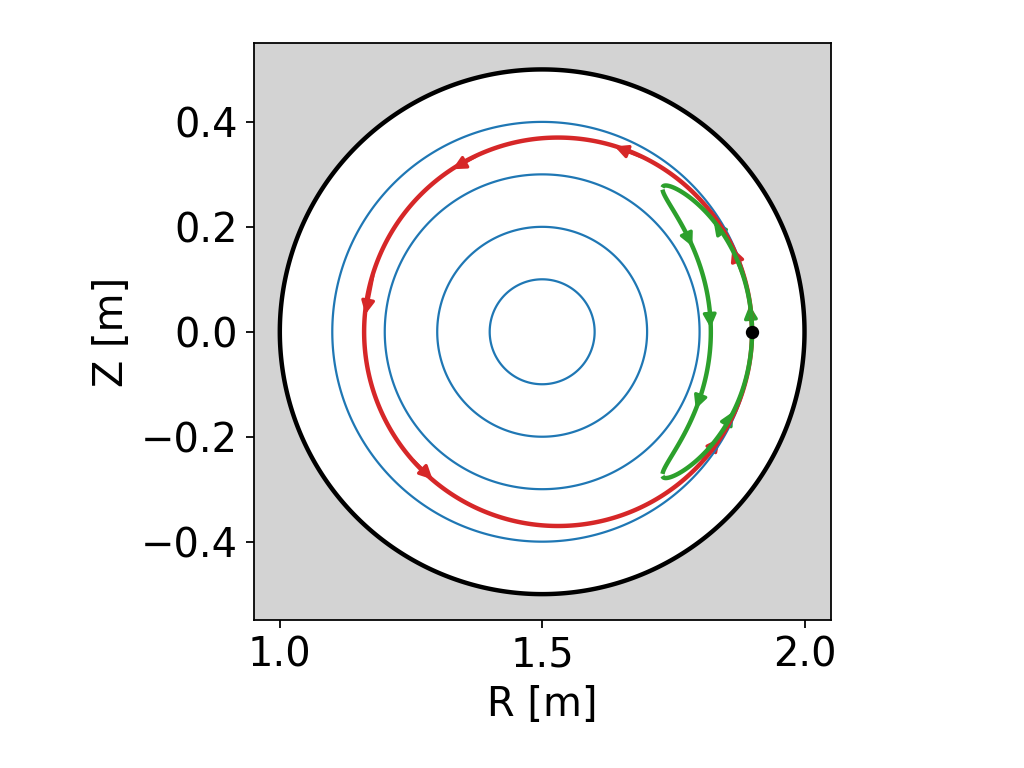}}
\subfigure[]{\includegraphics[scale=0.5]{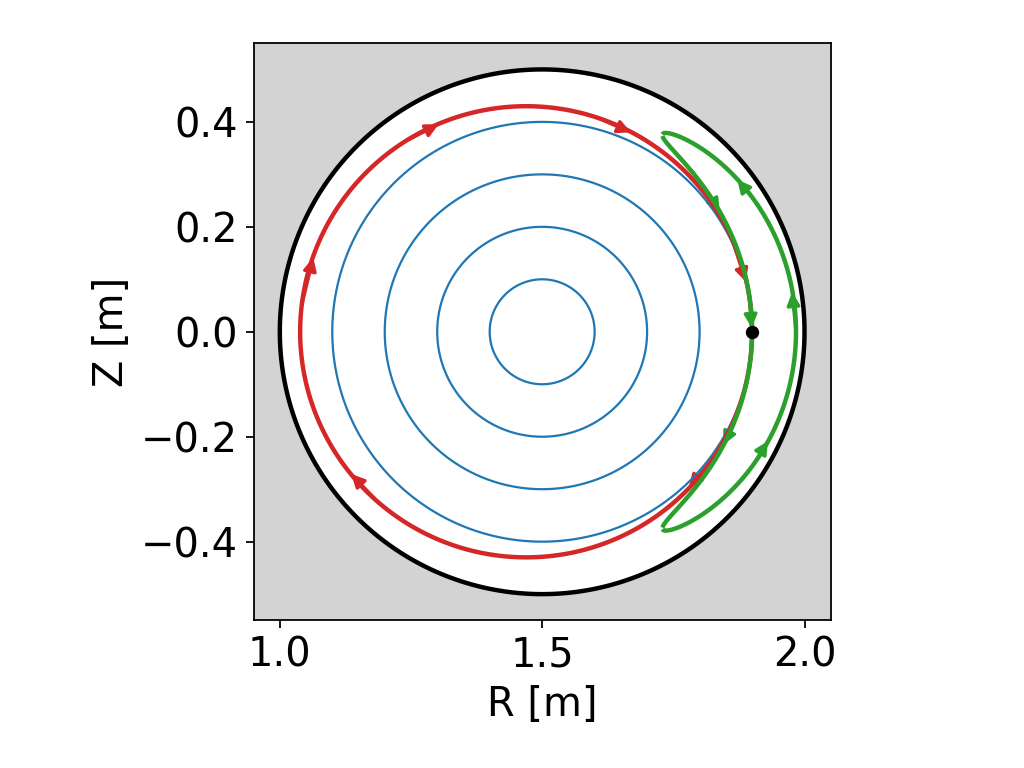}}
\par\end{centering}
\caption{Circular flux surface geometry used in this work and example collisionless ion orbits. Panel (a) shows co-current ions with a passing ion with initial pitch $\xi=0.8$ (red curve), and a trapped ion with initial pitch $\xi=0.3$ (green curve). Panel (b) shows counter-current ions with a passing ion with initial pitch $\xi=-0.8$ (red curve), and a trapped ion with initial pitch $\xi=-0.3$ (green curve). Flux surface contours are shown in blue. The black marker indicates the initial position of the ions, and arrows indicate their direction. The grad-B drift points downward in our example geometry. We assumed a deuterium ion with $20\;\text{keV}$, a tokamak with a minor radius of $0.5\;\text{m}$, inverse aspect ratio $a/R_0=1/3$, constant $q$-profile of $q=2$, and an on-axis magnetic field strength of $B_0=2\;\text{T}$.}
\label{fig:ET0}
\end{figure}

Before discussing solutions to the adjoint drift kinetic equation, it will be useful to describe the magnetic geometry employed. A plot of the circular flux surface geometry assumed, together with four example ion orbits, are shown in Fig. \ref{fig:ET0}. Here, the blue contours indicate magnetic flux surfaces, and the red and green curves indicate example passing and trapped ion orbits, respectively. The toroidal magnetic field and current are assumed to come out of the page, with the poloidal magnetic field in the counter-clockwise direction. Passing ions with a positive pitch (co-current), will thus rotate in the counter-clockwise direction when their motion is projected onto the poloidal plane as they propagate along a magnetic field line, with a grad-B drift vertically downward. Ions with a negative pitch (counter-current) will move in a clockwise sense when moving along field lines, but will still have a vertically downward grad-B drift. From Fig. \ref{fig:ET0}, it is apparent that co-current ions [Fig. \ref{fig:ET0}(a)] born on the weak field side will tend to be better confined compared to counter-current ions [Fig. \ref{fig:ET0}(b)] born at the same location, due to the magnetic drifts taking the counter-current ions closer to the spatial boundary. Furthermore, a counter-current passing ion when scattered into the trapped region will move onto an orbit that takes it closer to the plasma edge, as evident from Fig. \ref{fig:ET0}(b), such that these ions will be more susceptible to being lost.

\section{\label{sec:ET}Mean Escape Time}

\begin{figure}
\begin{centering}
\subfigure[]{\includegraphics[scale=0.5]{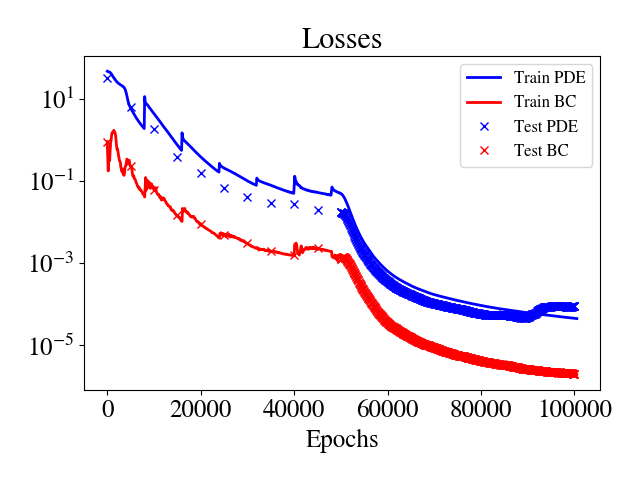}}
\subfigure[]{\includegraphics[scale=0.5]{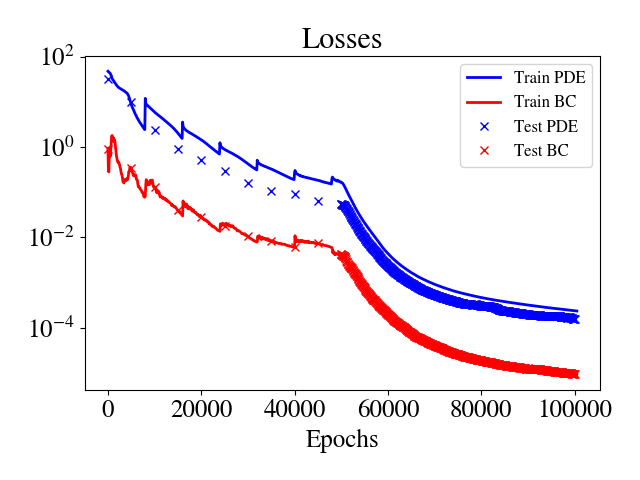}}
\par\end{centering}
\caption{Loss histories for 20 keV (panel a.) and 50 keV (panel b.) ions. Solid lines indicate the training loss whereas `x' markers indicate the test loss. Six million training points were used and two million test points. The same seed for the pseudorandom number generator was used in both cases.}
\label{fig:ET1}
\end{figure}

\subsection{Training Details of the PINN}

This section will describe solutions for the mean escape time of a fast ion population at 20 and 50 keV utilizing the PINN implementation described in Sec. \ref{sec:PINN}, along with a comparison to solutions evaluated with JONTA. Loss histories for ions with 20 and 50 keV are shown in Fig. \ref{fig:ET1}. For these cases, we assumed a midsize tokamak with a minor radius of $a=0.5\;\text{m}$, aspect ratio $R_0/a = 3$, and deuterium fast ions. The background deuterium plasma is assumed to have a temperature of $1\;\text{keV}$, a density of $n_i = 10^{20}\;\text{m}^{-3}$, and the safety factor is taken to be $q=2$. A representative Coulomb logarithm of $\ln \Lambda = 15$ was used in the collision operator. Each of these quantities were taken to be constant across the plasma radius, though nontrivial spatial profiles could easily be introduced. A fully connected feedforward network with seven hidden layers each with a width of fifty neurons was used. A network of this size, together with six million training points used the majority of the 180 GB of VRAM present on the NVIDIA Blackwell 200 GPU used to train the network. While we have observed improvements in predictions of the PINN by increasing the size of the network, we found the present network size and number of training points offers a good balance in ensuring dense coverage over the ion phase space, while providing a sufficiently expressive network to approximate the solution. For a network of this size, the inference time on an M3 MacBook Pro was measured to be two microseconds per prediction. 

The SOAP optimizer was used for the first 50,000 epochs, with SSBroyden used for the remaining 50,000 epochs. Six million training points were used with two million uniformly distributed test points applied. An adaptive residual based sampling method was employed during the SOAP phase of training~\cite{wu2023comprehensive}, such that training points are redistributed near regions with large residuals. From Fig. \ref{fig:ET1}, it is evident that SOAP drives a slow decay of the loss, with SSBroyden driving a sharper decline. The periodic spikes in the training loss during the SOAP phase are due to the training points being resampled every 8,000 epochs. As a result of training points being concentrated in regions with large residuals, the test loss is consistently lower than the training loss, except for the last 10,000 epochs of training for the case of 20 keV ions. This crossing of the test and training loss is due to a large residual forming at some isolated regions immediately adjacent to $r=a$, and will be discussed further below. We note that the SOAP optimizer being employed in this paper is able to efficiently train on a GPU, whereas the SciPy implementation of SSBroyden that we have employed primarily uses a CPU. As a result, the vast majority of the training time is spent during the SSBroyden phase of optimization.

\begin{figure}
\begin{centering}
\subfigure[]{\includegraphics[scale=0.33]{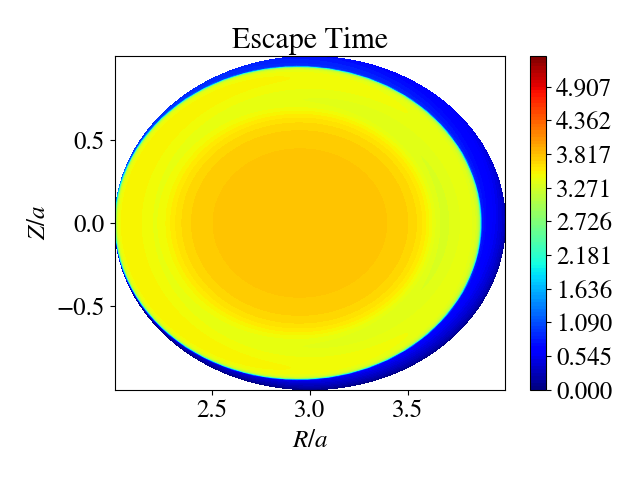}}
\subfigure[]{\includegraphics[scale=0.33]{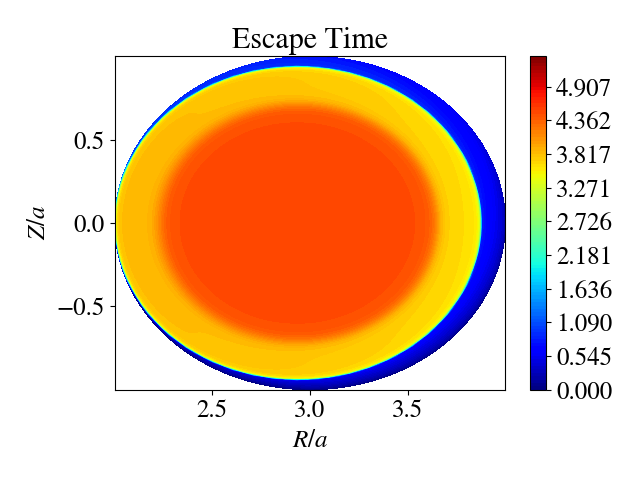}}
\subfigure[]{\includegraphics[scale=0.33]{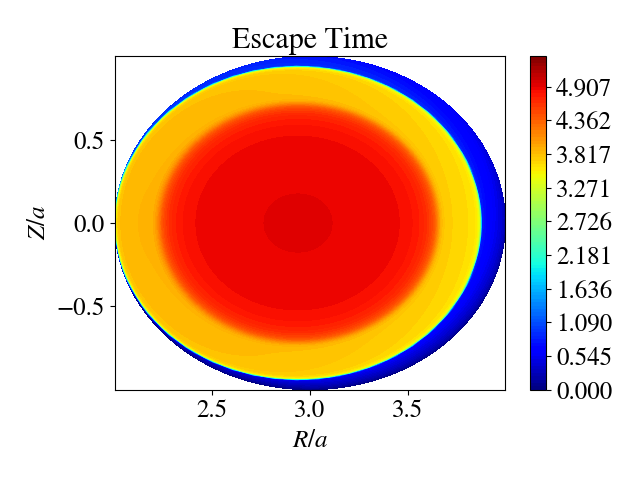}}
\subfigure[]{\includegraphics[scale=0.33]{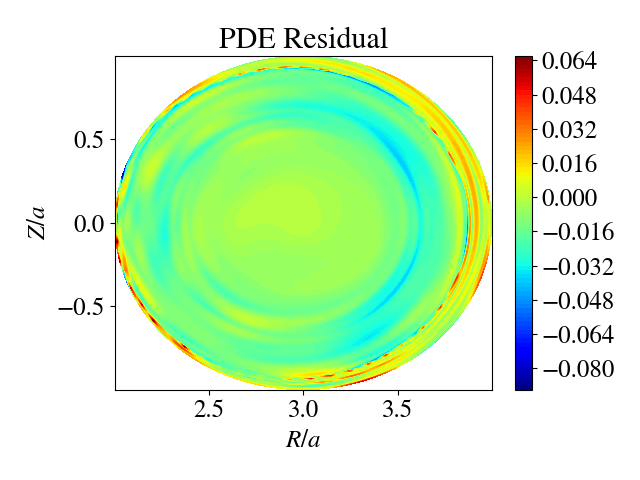}}
\subfigure[]{\includegraphics[scale=0.33]{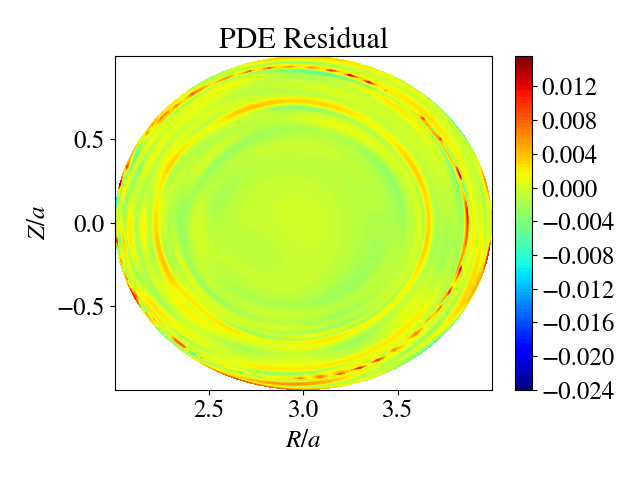}}
\subfigure[]{\includegraphics[scale=0.33]{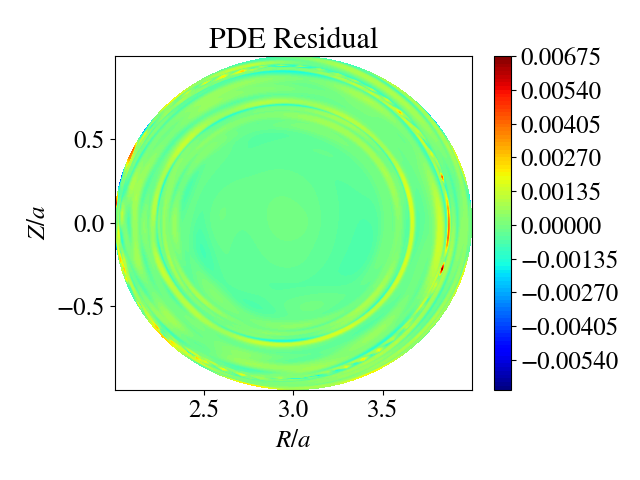}}
\par\end{centering}
\caption{(a) Mean escape time for an ion initialized in the counter-current direction with a pitch of $\xi = -0.8$, after different stages of training. The profile of the ion's escape time is shown in the panels (a), (b) and (c), with the residual indicated in panels (d), (e) and (f). The first column is after the SOAP phase of training (50,000 epochs), the second column is 10,000 into the SSBroyden phase (60,000 net epochs), and the final column is after 50,000 epochs of SSBroyden (100,000 epochs total). The quantity plotted is $\text{log}_{10}\left( 1+ T\right)$, yielding a $\text{log}_{10}$ scale for large values of $T$, but vanishing when $T=0$. Deuterium ions with $20\;\text{keV}$ were assumed.}
\label{fig:ET1sub1}
\end{figure}

\begin{figure}
\begin{centering}
\subfigure[]{\includegraphics[scale=0.5]{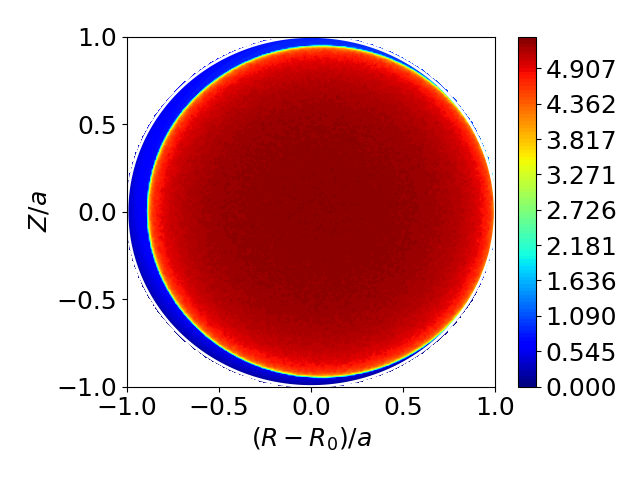}}
\subfigure[]{\includegraphics[scale=0.5]{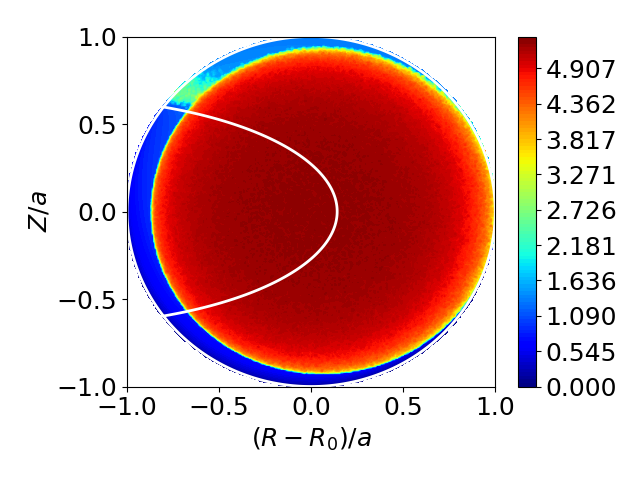}}
\subfigure[]{\includegraphics[scale=0.5]{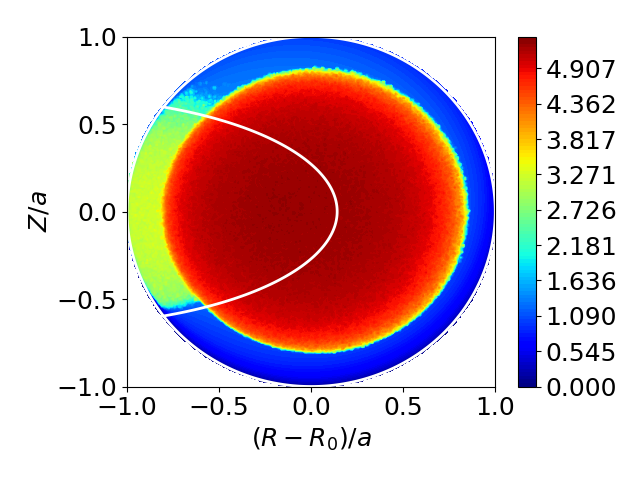}}
\subfigure[]{\includegraphics[scale=0.5]{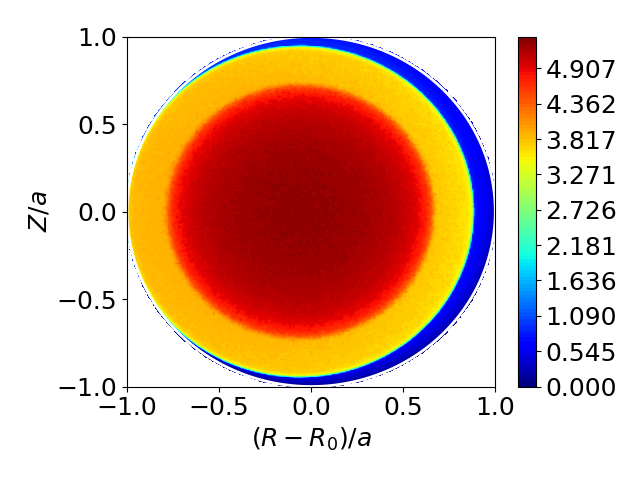}}
\par\end{centering}
\caption{Slices of escape time in units of $\log_{10}(1+T)$ for different initial pitch values and an energy of $20\;\text{keV}$ computed from the JONTA code. Panel (a) is for $\xi=0.8$, (b) is for $\xi=0.3$, (c) is for $\xi=-0.3$, and (d) is for $\xi=-0.8$. Ten million markers were used, where the particles were integrated for $t_{final} = 2\times 10^6$.
}
\label{fig:ET6}
\end{figure}

Considering the evolution of the predicted solution during training, Fig. \ref{fig:ET1sub1}(a) shows the predicted mean escape time for a 20 keV ion after the SOAP optimization phase, 10,000 epochs into the SSBroyden phase [Fig \ref{fig:ET1sub1}(b)], and the final predicted mean escape time [Fig \ref{fig:ET1sub1}(c)]. A reference solution from JONTA is shown in Fig. \ref{fig:ET6}(d) below for comparison. It is evident that SOAP is able to approximately capture the structure of the solution near the tokamak edge, but fails to recover the long mean escape time of ions initially located at small minor radii. This is due to the large discrepancy in time scales between ions that begin on loss orbits, and are thus lost on a transit time scale, and those initially located deep in the plasma interior. These latter ions must collisionally diffuse out of the plasma, and are thus lost on a much longer time scale, roughly $10^5$ slower for the present example. After 10,000 epochs with SSBroyden, the training and test loss of the residual and boundary terms drop further, and it is evident from Fig. \ref{fig:ET1sub1} that the relatively long confinement time of ions initially located in the interior is better captured, though its value is still substantially under predicted. Finally, after 50,000 epochs of SSBroyden, the predicted escape time of ions initially located in the interior slowly increases, though the final value is still below that predicted by the particle-based code JONTA. It has been verified that additional training epochs with SSBroyden leads to a more accurate recovery of the mean escape time of the best confined ion orbits, though as described below, begins to overfit the solution near the plasma edge at isolated locations.

\begin{figure}
\begin{centering}
\includegraphics[scale=0.5]{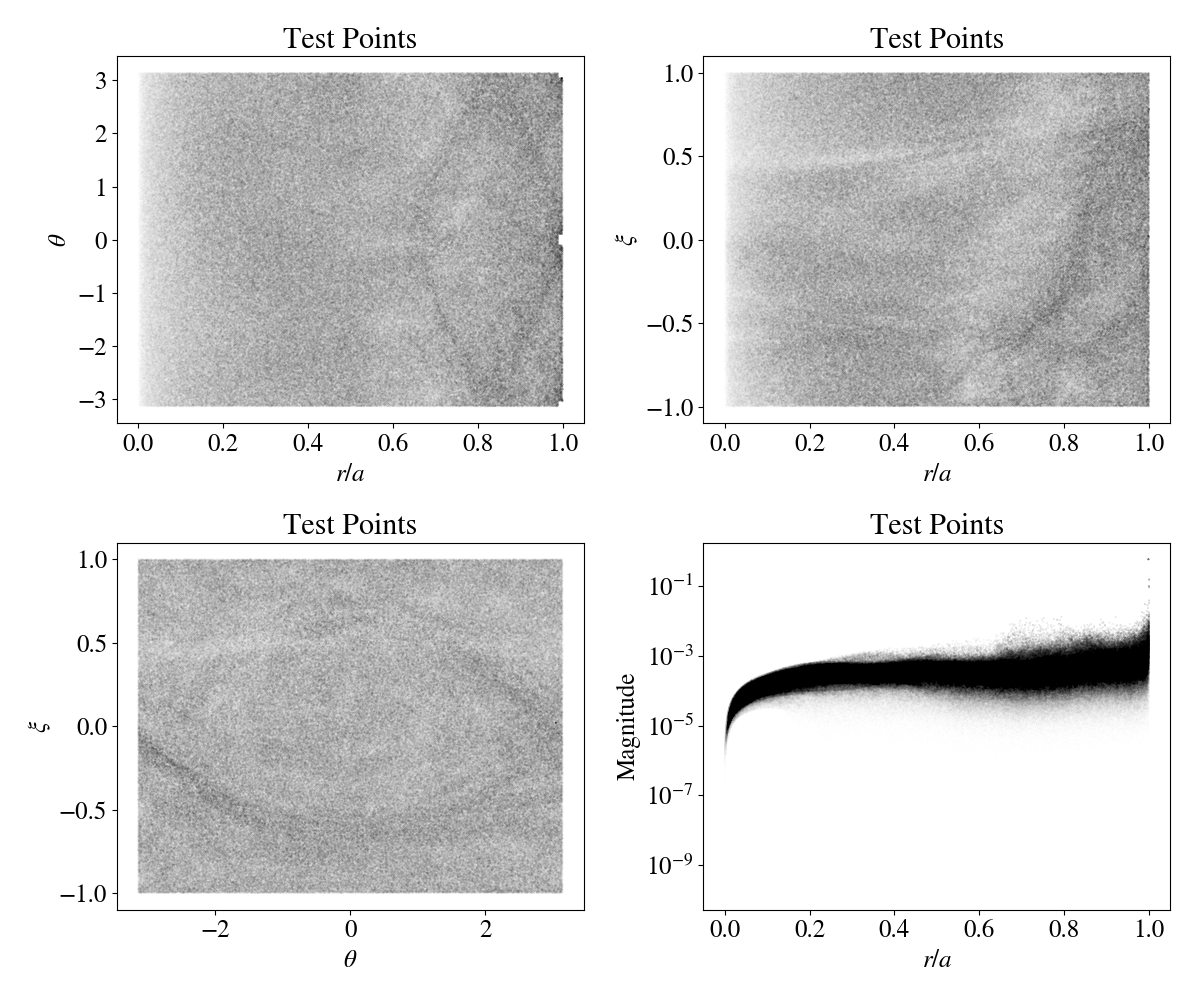}
\par\end{centering}
\caption{Test loss distribution for 20 keV ions. A uniform random distribution with one million points was used. The size of the markers is proportional to the magnitude of the residual.}
\label{fig:ET2}
\end{figure}

Turning to the distribution of training losses, two dimensional projections of the resulting test error are shown in Fig. \ref{fig:ET2}, with the magnitude of the markers set to be proportional to the magnitude of the residual. Hence, darker regions represent regions with the largest residuals. From Fig. \ref{fig:ET2}, the outer thirty percent of the tokamak contained the largest losses. This follows due to this region containing direct loss orbits, whose detailed form can be difficult to resolve. It is also evident from Fig. \ref{fig:ET2}(d), where the magnitude of the residuals are plotted as a function of the minor radius, that for $r\approx a$ a small number of very large residuals are present, with the largest having a magnitude of $0.589$ located at $r/a \approx 0.998$ and $\theta \approx 3.041$. These large residuals at the very edge of the plasma are responsible for the increase in the test loss during the last 10,000 epochs of training, but as discussed further below, do not substantially impact the accuracy of the solution throughout the majority of the plasma volume.
Finally, regions with $r/a > 0.99$ and poloidal angles on either the outboard ($-\Delta \theta < \theta < \Delta \theta$) or inboard ($\theta > -\pi + \Delta \theta$ and $\theta < \pi - \Delta \theta$) midplanes were set to zero, leading to empty regions in Fig. \ref{fig:ET2}(a). For this example we set $\Delta \theta = 0.1$. 
As described further below, these regions contain discontinuities in the solution arising from changes in boundary conditions between the lower and upper half of the tokamak, which cannot be precisely resolved. By removing training points from this region, this allows the PINN to find a solution that smoothly transitions across these narrow regions. We note that the largest residuals in the test loss are located at poloidal angles just outside of these regions, and are thus due to the sharp variation of the solution between the upper and lower regions of the tokamak near the plasma edge [see Fig. \ref{fig:ET5}(a) below].

\subsection{Phase Space Structure of the Mean Escape Time}

\begin{figure}
\begin{centering}
\subfigure[]{\includegraphics[scale=0.5]{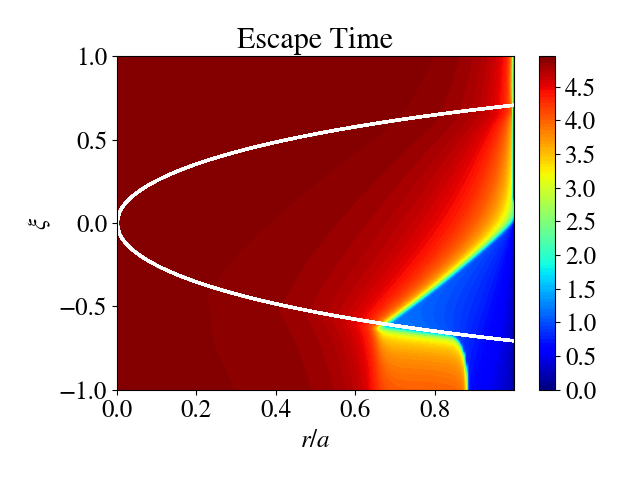}}
\subfigure[]{\includegraphics[scale=0.5]{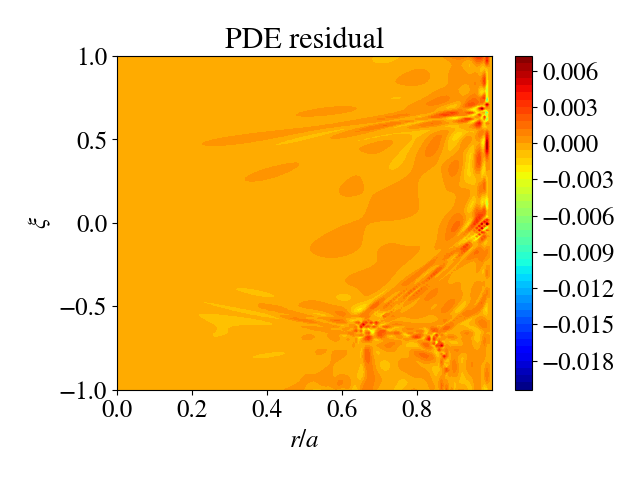}}
\par\end{centering}
\caption{(a) Escape time for an ion initially located at $\theta = 0$. The quantity plotted is $\text{log}_{10}\left( 1+ T\right)$, yielding a $\text{log}_{10}$ scale for large values of $T$, but vanishing when $T=0$.  The white curve indicates the trapped-passing boundary. (b) Residual of Eq. (\ref{eq:IADK1}). A deuterium ion with $20\;\text{keV}$ was assumed.}
\label{fig:ET3}
\end{figure}

The phase space distribution of the fast ion mean escape time can be conveniently illustrated by taking slices in the three-dimensional $\left( R, Z, \xi\right)$ space. 
The mean escape time of ions born with $\theta = 0$ (weak field side) is shown in Fig. \ref{fig:ET3} on a $\log_{10}$ scale. Here, counter-current ions ($\xi < 0$) born in the edge of the plasma are lost quickly due to direct orbit loss, with both co and counter current ions initially located deeper in the plasma having much longer mean escape times. Further noting that the white contour indicates the trapped-passing boundary, it is evident that co-current passing ions (positive pitches above the white contour) have far better confinement in comparison to counter-current ions. Similarly, in the trapped region, ions that begin with $\xi > 0$ are much better confined compared to ions with $\xi < 0$. Comparing the left and right panels of Fig. \ref{fig:ET0}, this difference in confinement results from trapped ions born with $\xi < 0$ at $\theta = 0$ having orbits that extend out to larger radii [Fig. \ref{fig:ET0}(b)], in contrast to trapped ions born with $\xi>0$ [Fig. \ref{fig:ET0}(a)]. The ions with the worst confinement are those just inside the trapped-passing boundary and $\xi < 0$, which are expected to have the largest banana width, and thus be most susceptible to orbit loss. 


\begin{figure}
\begin{centering}
\subfigure[]{\includegraphics[scale=0.5]{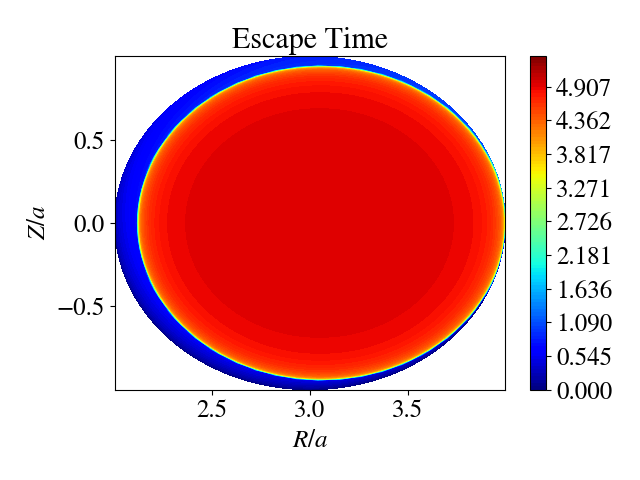}}
\subfigure[]{\includegraphics[scale=0.5]{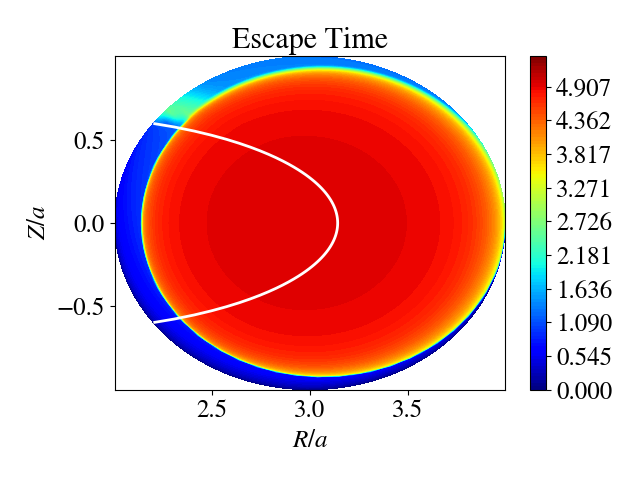}}
\subfigure[]{\includegraphics[scale=0.5]{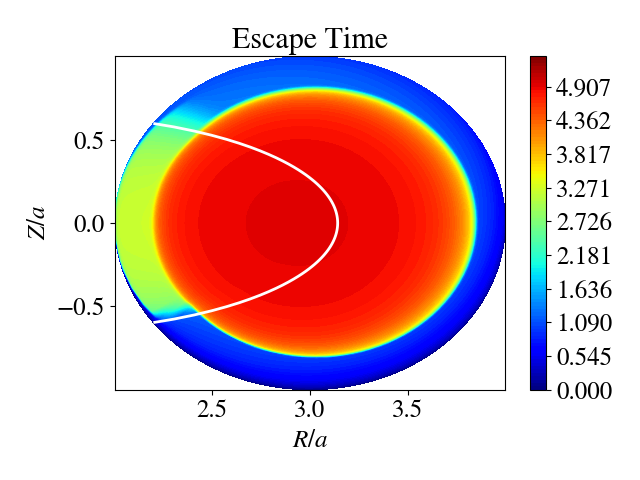}}
\subfigure[]{\includegraphics[scale=0.5]{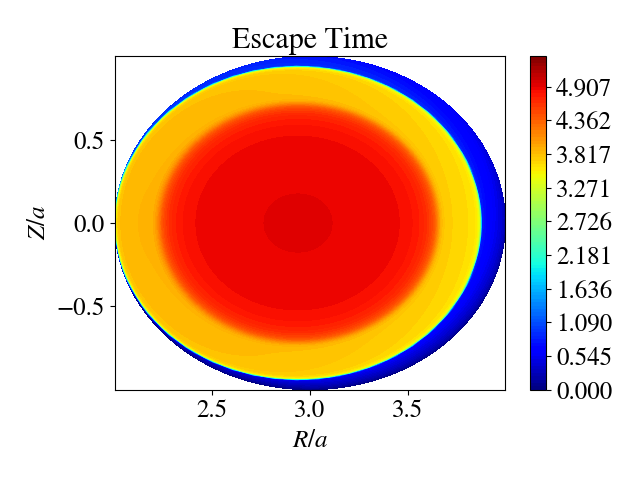}}
\par\end{centering}
\caption{Slices of escape time in units of $\log_{10}(1+T)$ for different initial pitch values. Panel (a) is for $\xi=0.8$, (b) is for $\xi=0.3$, (c) is for $\xi=-0.3$, and (d) is for $\xi=-0.8$. The solid white curve indicates the location of the trapped-passing boundary, where ions born with major radii $R$ to the right of the white curve are trapped.}
\label{fig:ET4}
\end{figure}

Figure \ref{fig:ET4} shows slices of major radius $R$ and vertical height $Z$ for different pitches. These slices allow for the difference in confinement of co and counter current ions to be investigated more clearly. Considering passing co-current ions ($\xi=0.8$), these ions will propagate in the counter-clockwise direction in Fig. \ref{fig:ET4}(a). Noting that ion orbits drift downward, this will result in ions that are initially on the outboard side not being on direct loss orbits, in contrast to those born on the inboard side. This leads to a notable in-out asymmetry of the fast ion population, as evident in Fig. \ref{fig:ET4}(a). This trend is reversed for passing ions that begin in the counter-current direction [Fig. \ref{fig:ET4}(d)]. A more subtle feature is that counter-current ions that are not on direct loss orbits are more likely to be lost in comparison to co-current ions. This is evident by the orange region in Fig. \ref{fig:ET4}(d). The relatively poor confinement of these counter-current ions is due to the collisional trapping of these ions resulting in the orbit exploring larger minor radii [compare the red and green curves in Fig. \ref{fig:ET0}(b)]. Hence, passing counter-current ions that are within a banana width of the plasma edge are far more susceptible to being lost to the plasma compared to passing co-current ions. 


\begin{figure}
\begin{centering}
\subfigure[]{\includegraphics[scale=0.5]{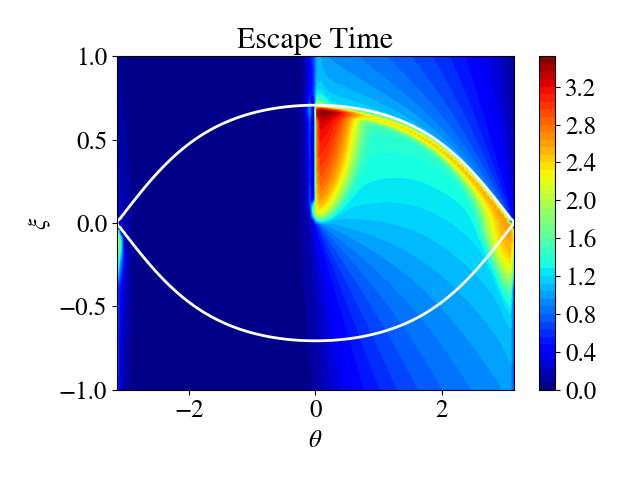}}
\subfigure[]{\includegraphics[scale=0.5]{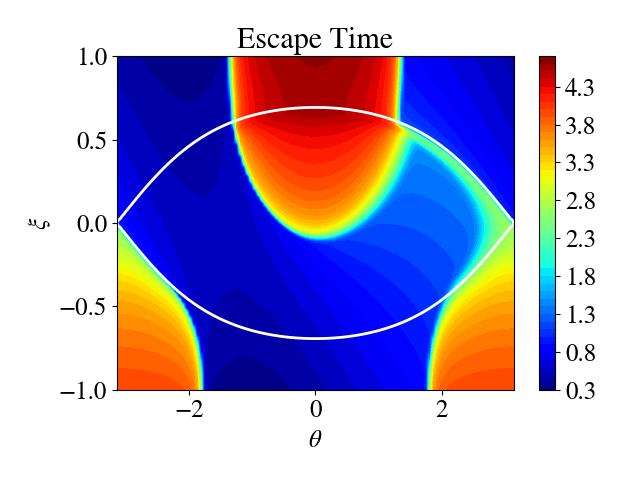}}
\subfigure[]{\includegraphics[scale=0.5]{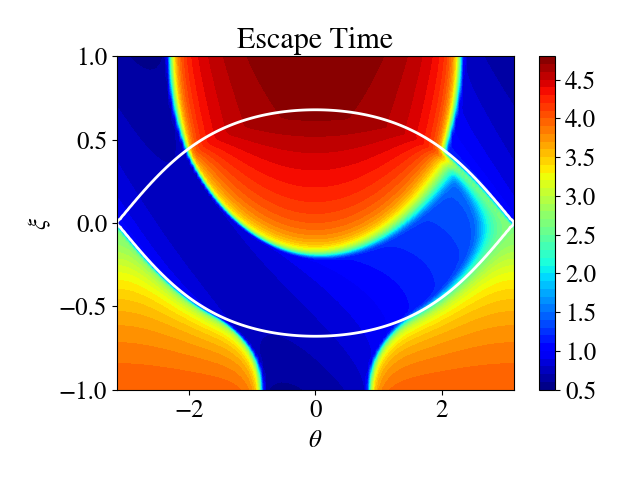}}
\subfigure[]{\includegraphics[scale=0.5]{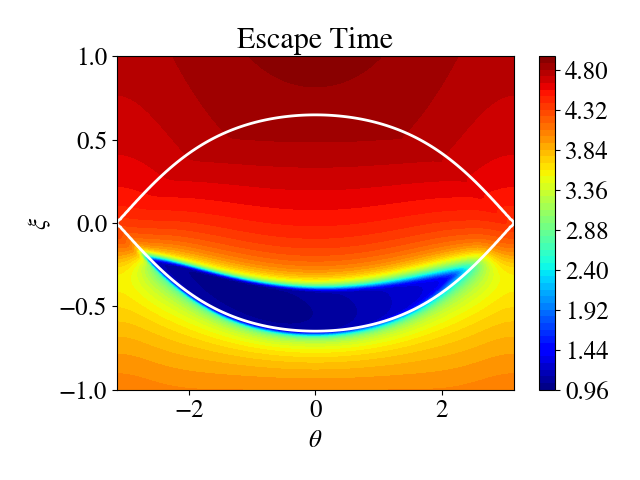}}
\par\end{centering}
\caption{Slices of escape time in units of $\log_{10}(1+T)$ for different initial radii. Panel (a) is for $r/a=0.999$, (b) is for $r/a=0.95$, (c) is for $r/a=0.9$, and (d) is for $r/a=0.8$. The while curves are the trapped-passing boundary.}
\label{fig:ET5}
\end{figure}

Cross-cuts of the pitch and poloidal angle distribution of loss times at different radii are shown in Fig. \ref{fig:ET5}. Here, when approaching $r/a=1$ a discontinuity appears in the solution. This is evident in Fig. \ref{fig:ET5}(a), where for $\theta \approx 0$ and $\theta \approx \pm \pi$, the solution changes rapidly when traversing these discontinuous regions. The origin of this discontinuity is that for ions in the lower half of the tokamak (i.e. $-\pi < \theta < 0$) and $r/a=1$, they are lost immediately. In contrast, ions located just above $\theta = 0$ or $\theta = \pm \pi$ are often not on direct loss orbits, and thus can be confined for a long time. For example, co-current ions just above $\theta = 0$, will drift inward, and thus have a finite escape time, whereas those just below $\theta = 0$ are lost immediately to the boundary. An analogous discontinuity in the solution is present for $\theta = \pm \pi$ for counter-current ions. To relax these discontinuities, we have removed all training and test points from these regions, such that the PINN is able to approximate these discontinuities with sharply varying, but continuous solutions, as evident in Fig. \ref{fig:ET5}(a). Moving radially away from the $r/a=1$ boundary, panels (b)-(d) of Fig. \ref{fig:ET5} show the solution at radial locations between $r/a = 0.95$ and $r/a=0.8$, indicating vastly different mean escape times of ions initially located at different pitch and poloidal angles. In particular, the phase space region of prompt ion loss, which initially occupies roughly half of the $\left( \xi, \theta \right)$ domain for $r/a\approx 1$, shrinks to a narrow sliver just inside the trapped-passing boundary for $r/a=0.8$. This is indicated in Fig. \ref{fig:ET5}(d), where the solid white curve indicates the trapped-passing region.


\subsection{Comparison with Monte Carlo Solutions}

The mean escape time can also be computed by the particle-based solver JONTA described in Sec. \ref{sec:PBDK}. Here, marker particles are initialized randomly across the spatial domain at a given pitch. Once a particle's minor radius exceeds $r_{wall}$, its escape time and location are saved. By pushing a large number of marker particles, the mean escape time can be computed as a function of the ion's initial $R$ and $Z$. Noting the conceptual simplicity of this approach, we will use it to validate the predictions of the PINN. Figure \ref{fig:ET6} shows four slices in the $\left( R, Z \right)$ plane at constant pitch computed from JONTA. Here, ten million marker particles are randomly distributed at a given pitch, with their escape times recorded after they escape from the domain. These escape times are then binned, allowing for the mean escape time to be computed over the $\left( R, Z \right)$ plane for a given pitch. The particles were evolved until $t_{final} = 2\times 10^6$, a time at which nearly all of the original particles remained in the domain. Four cross cuts are shown in Fig. \ref{fig:ET6}. These four cross cuts are for the same physics parameters and pitches as Fig. \ref{fig:ET4} above. Good agreement between the mean escape time predicted by JONTA and the PINN are evident. In particular, both approaches are able to recover the rapid escape of ions that are initiated on direct loss orbits (blue regions), ions lost due to collisional trapping/detrapping directly onto a loss orbit (yellow and green regions), and ions initially located near the magnetic origin that must slowly diffuse out (red regions). The JONTA simulations also reveal a limitation of our implementation of a PINN. In particular, while the PINN is able to distinguish the complex edge structure of confined versus unconfined orbits, with excellent agreement with predictions of JONTA, it struggles to accurately capture the very long mean escape time for ions initially located deep inside the plasma. Specifically, the dark red region evident in Fig. \ref{fig:ET6} of the JONTA data, is not present in the PINN's solution (see Fig. \ref{fig:ET4}), indicating roughly a factor of two and a half discrepancy in predicted confinement for these 20 keV well confined ions. 
The reason for this quantitative inaccuracy is that the PINN struggles to simultaneously resolve the fast direct orbit dynamics together with the slow collisional transport of the core ions, though as indicated in Fig. \ref{fig:ET1sub1}, this discrepancy slowly shrinks as the model is trained. The resolution of this disparate physics will require (i) the inclusion of data in training the PINN, e.g. using data taken directly from JONTA for example, (ii) allowing the PINN to train for more epochs which has been verified to slowly reduce the discrepancy, or (iii) modifications to the training of the PINN. These extensions will be discussed further below.


\begin{figure}
\begin{centering}
\subfigure[]{\includegraphics[scale=0.24]{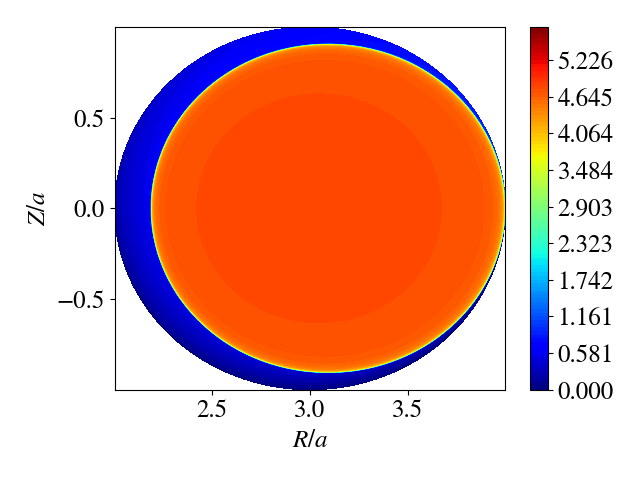}}
\subfigure[]{\includegraphics[scale=0.24]{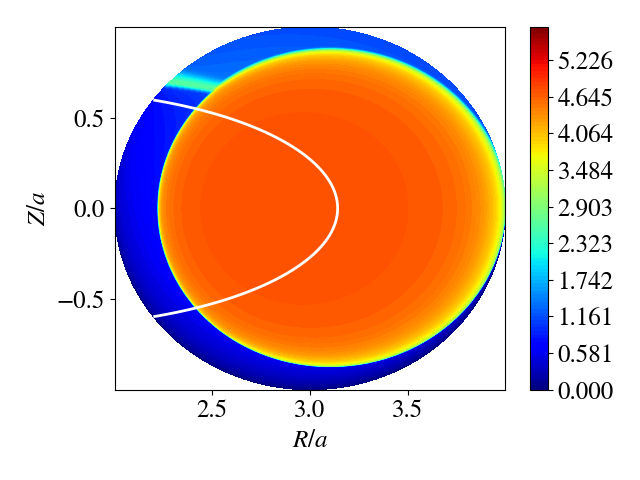}}
\subfigure[]{\includegraphics[scale=0.24]{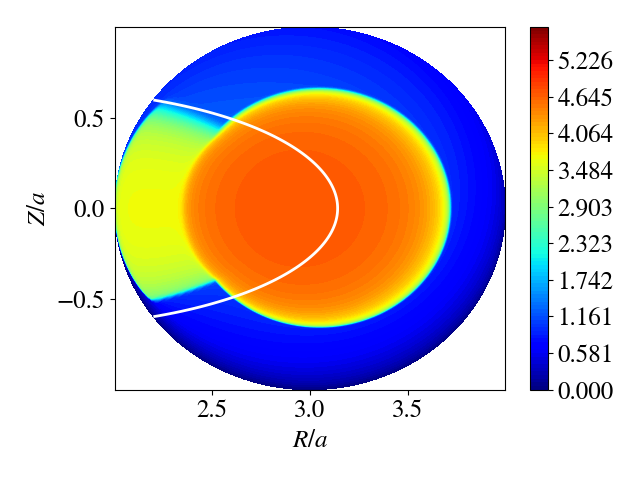}}
\subfigure[]{\includegraphics[scale=0.24]{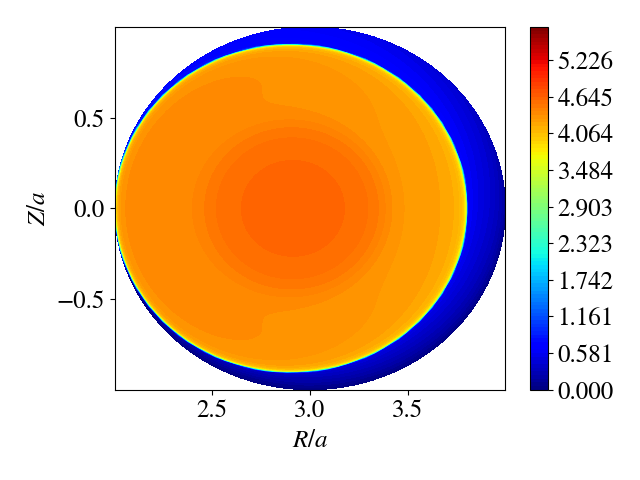}}
\subfigure[]{\includegraphics[scale=0.24]{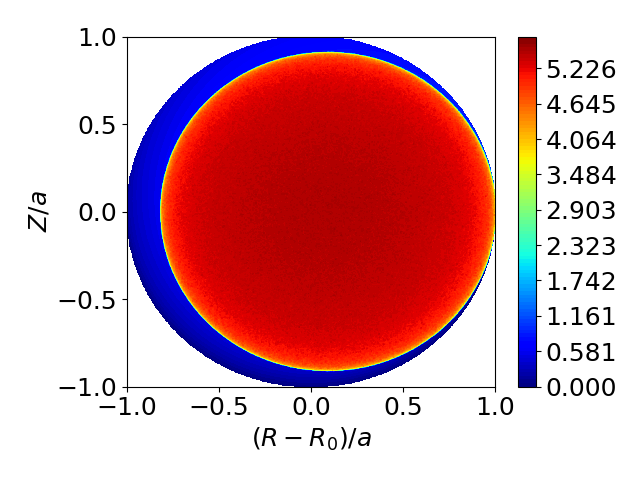}}
\subfigure[]{\includegraphics[scale=0.24]{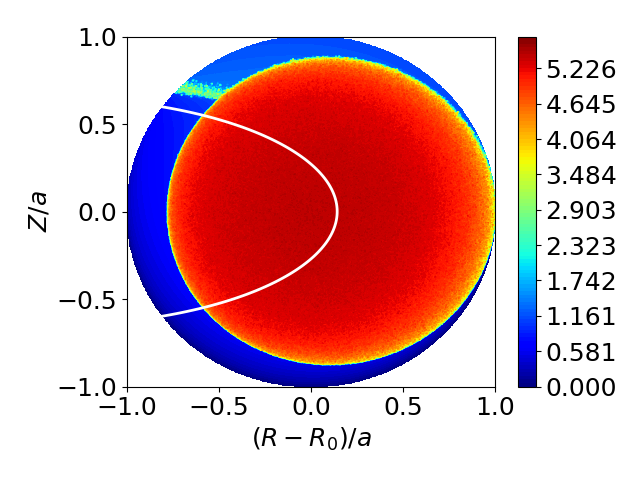}}
\subfigure[]{\includegraphics[scale=0.24]{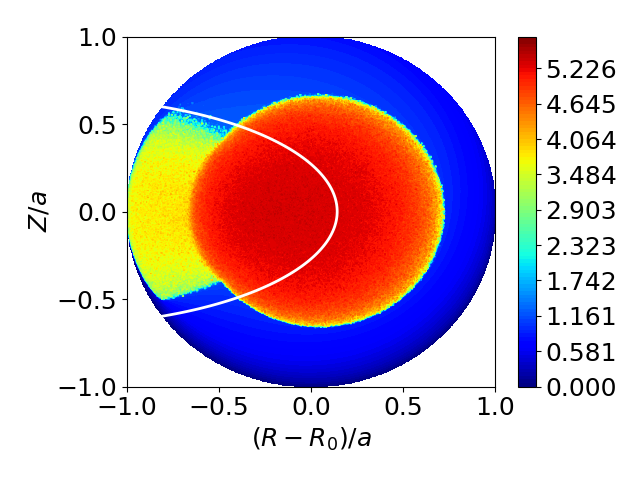}}
\subfigure[]{\includegraphics[scale=0.24]{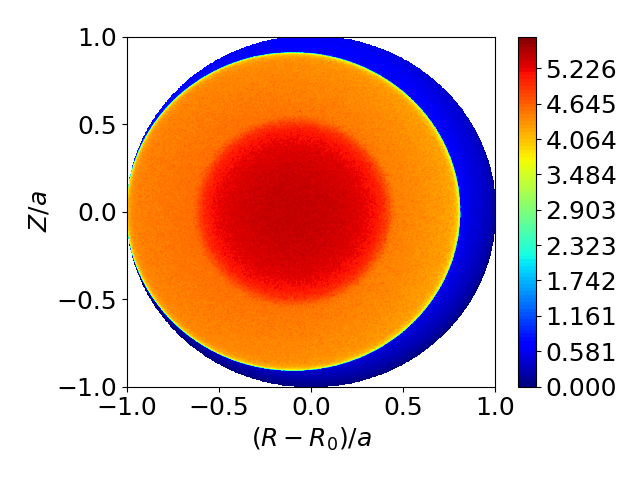}}
\par\end{centering}
\caption{Comparison of escape time in units of $\log_{10}(1+T)$ predictions from the PINN (top row) and JONTA (bottom row). Ions were assumed to have an initial energy of $50\;\text{keV}$. Panels (a,e) are for $\xi=0.8$, (b,f) are for $\xi=0.3$, (c,g) are for $\xi=-0.3$, and (d,h) are for $\xi=-0.8$. Ten million markers were used in the JONTA simulations, where the particles were integrated for $t_{final} = 10^7$.
}
\label{fig:ET7}
\end{figure}

Now considering an energetic ion population with $50\;\text{keV}$, similar physical trends are evident, but the larger ion orbits lead to a larger number of ions beginning on direct loss orbits (see Fig. \ref{fig:ET7}). In addition, the higher energy of the particles leads to a lower collision time, and thus energetic ions initially located in the core have a longer escape time. This latter property results in the problem exhibiting greater stiffness, with a faster bounce time but slower collision time, resulting in the PINN struggling to accurately resolve the escape time of the best confined orbits. This is evident in Fig. \ref{fig:ET7}, where cross sections in the $\left( R, Z \right)$ plane computed both with JONTA and the PINN are compared. While the PINN fails to capture the precise mean escape time of the best confined particles, it does succeed in accurately capturing the phase space structure of the mean escape time across the edge region, where ions have the shortest mean escape times. It thus acts as an effective tool for delineating well confined regions of phase space, from those where ions are directly lost, or lost after a small number of collisional scattering events. We note that while for the case of $20\;\text{keV}$ ions a factor of roughly two and a half difference in the mean escape time was present for the best confined ions, for $50\;\text{keV}$ roughly a factor of five difference in the prediction of the best confined orbits is evident. This larger deviation in the prediction of the best confined ions is due to the increased time scale separation between the collisionless ion orbits and the slow collision time. As discussed further in Sec. \ref{sec:C}, while further training or hyperparameter tuning can likely reduce these discrepancies, a robust means of improving the predictions of the PINN will be the incorporation of small quantities of data.

\section{\label{sec:C}Conclusions}


The inhomogeneous adjoint of the drift kinetic equation was used to evaluate the mean escape time of energetic ion populations. This quantity, computed in the limit of an axisymmetric tokamak plasma, captures fast ion losses due to direct orbit loss and collisional transport, thus providing a metric for energetic particle confinement. A PINN was used to evaluate the phase space dependence of the mean escape time, a challenging task due to the strong time scale separation between the fast bounce time of energetic ions, and their slow collision rate. While the offline training time of the PINN is substantial, the online inference time is rapid, typically microseconds per prediction. Although the PINN failed to quantitatively capture the mean escape time of the best confined ion orbits, it is able to robustly recover the phase space structure of the mean escape time, and thus acts as a means of distinguishing well confined orbits and phase space regions of prompt ion loss. We also note that the present analysis did not include fast ion slowing down, where the slowing down time for the 20 and 50 keV ions considered in this work would be far shorter than the mean escape times computed in this work for the best confined orbits. 

For this proof-of-principle demonstration of the proposed framework we considered several simplifications to the complete system. Perhaps the primary simplifications were the assumption of (i) an axisymmetric magnetic geometry, (ii) a simplified collision operator, and (iii) idealized plasma profiles and shaping. Regarding item (i), since PINNs to date have not been used to solve the drift kinetic equation, beginning with an axisymmetry geometry was a logical starting point for assessing this approach's strengths and weaknesses. We anticipate extending the present approach in future work to 3D magnetic geometry~\cite{law2022accelerating}, though such an extension will likely require incorporating particle data into the training of the neural network. The particle-based solver JONTA will provide a convenient tool through which such data can be generated, where by providing an accurate calculation of the mean escape time at specific phase space locations, these may be included in the first term of the loss function defined by Eq. (\ref{eq:PINN1}). By providing high fidelity data points at internal locations, for example in regions with large residuals or for the best confined ion orbits where the present implementation of a PINN has struggled, this will likely expedite the training of the PINN while providing a systematic means of improving its accuracy. Our aim will be to leverage data as a means of both treating more challenging scenarios, such as non-axisymmetric geometry, along with facilitating the learning of a parametric solution to the drift kinetic equation, enabling the mean escape time to be evaluated for a broad range of plasma conditions.

An additional simplification was the use of a Lorentz collision operator, which does not account for the slowing down of energetic particles, a quantity of critical interest when assessing the amount of energy deposited by energetic ions into the bulk plasma. Despite this limitation, including Lorentz collisions allows for the collisional transport of fast ions to be assessed. Thus, while not providing an accurate indication of the mean escape time for well confined energetic particles, since such particles will likely slow down substantially during their long confinement time, it does provide a means of rapidly assessing particles on direct loss orbits, or those that are directly lost after being collisionally scattered into or out of the trapped particle region. The predictions of the present work will thus be most relevant to the outer region of the tokamak, where energetic particles are lost on relatively short time scales. Finally, homogeneous plasma profiles and a circularly shaped plasma were assumed in the present work. This was done for convenience, where we anticipate including more realistic plasma profiles to not pose a substantial challenge. Furthermore, since PINNs are mesh free, and are thus well suited to treating shaped magnetic geometry~\cite{van1995neural, kaltsas2022neural, jang2024grad}, we do not expect the treatment of shaped axisymmetric geometry to pose a major hurdle. Similarly, strong radial electric fields that emerge in the pedestal region~\cite{brzozowski2019geometric} along with realistic equilibrium density and temperature profiles could easily be incorporated.

Due to the strong time scale separation, evaluating fast ion transport is computationally intensive, creating a bottleneck in many optimization workflows. For the JONTA simulations presented in this paper, the shortest simulation required one and a half hours on a single NVIDIA Blackwell 200 GPU, with the longest requiring slightly over a day. Noting that each JONTA simulation was for a fixed value of pitch, computing four distinct pitches (Figs. \ref{fig:ET6} and \ref{fig:ET7}) thus required several hours to days of computing time. A PINN, which encompasses all values of pitch, and thus only a single PINN must be trained, required approximately three days to train, where approximate results obtained after the SOAP phase of training typically take a few hours. Both approaches are thus computationally intensive, where while the PINN is able to recover smooth predictions of the mean escape time, it is unable to accurately compute the best confined ions. JONTA, in contrast, requires roughly ten million markers to obtain a relatively smooth, well resolved solution. We note that while the PINN also used a single NVIDIA Blackwell 200 GPU, the implementation of the SSBroyden optimizer does not make efficient use of the GPU, and thus was mostly trained using a CPU, which was the primary bottleneck in the training of the PINN.


We also note that while the present paper has focused on the mean escape time of energetic ions, several straightforward extensions are possible. One such extension would be to treat a broader range of fast ion metrics. Metrics of interest include the probability that an ion impacts a specific region of the wall from a given initial phase space location~\cite{sarkimaki2020efficient, salewski2011velocity}, or the slowing down probability, which corresponds to the probability that an ion slows down to the thermal energy. Both of these fast ion metrics can be straightforwardly constructed from the steady state homogeneous adjoint equation by applying appropriate boundary conditions. The evaluation of this broader class of fast ion metrics will be left for future work. Finally, while we have considered energetic ions in the present formulation, extensions to the relativistic drift kinetic equation describing runaway electron (RE) evolution and transport can be developed following an analogous approach. Such extensions will allow for `neoclassical' corrections to RE generation rates to be inferred~\cite{Rosenbluth:1997, mcdevitt2019runaway, arnaud2024impact}, or account for spatial transport due to processes such as the Ware pinch~\cite{nilsson2015trapped, mcdevitt2018spatial}. While we anticipate that the extremely short bounce time of REs compared to their long collision times will pose a substantial challenge compared to the purely momentum space treatment employed in Refs. \cite{arnaud2024physics, mcdevitt2025primary}, we note that the impact of `neoclassical' effects on RE generation rates are most important near the critical energy for electrons to run away~\cite{mcdevitt2019runaway, arnaud2024impact}, an energy that often ranges between several keV to tens of keV during a tokamak disruption. At these modest energies, the time scale separation between the transit or bounce time of a RE and the collision time will be drastically reduced. The extension of the present framework to REs will be pursued in future work.

\begin{acknowledgements}

This work was supported by the Department of Energy, Office of Fusion Energy Sciences under awards DE-SC0024634 and DE-SC0024649, and the University of Florida Division of Sponsored Research Projects. The authors acknowledge the University of Florida Research Computing for providing computational resources that have contributed to the research results reported in this publication. 
  

\end{acknowledgements}

\newpage

\bibliographystyle{apsrev}
\bibliography{./ref}

\begin{thebibliography}{56}
\expandafter\ifx\csname natexlab\endcsname\relax\def\natexlab#1{#1}\fi
\expandafter\ifx\csname bibnamefont\endcsname\relax
  \def\bibnamefont#1{#1}\fi
\expandafter\ifx\csname bibfnamefont\endcsname\relax
  \def\bibfnamefont#1{#1}\fi
\expandafter\ifx\csname citenamefont\endcsname\relax
  \def\citenamefont#1{#1}\fi
\expandafter\ifx\csname url\endcsname\relax
  \def\url#1{\texttt{#1}}\fi
\expandafter\ifx\csname urlprefix\endcsname\relax\def\urlprefix{URL }\fi
\providecommand{\bibinfo}[2]{#2}
\providecommand{\eprint}[2][]{\url{#2}}

\bibitem[{\citenamefont{Hemsworth et~al.}(2008)\citenamefont{Hemsworth, Tanga,
  and Antoni}}]{hemsworth2008status}
\bibinfo{author}{\bibfnamefont{R.}~\bibnamefont{Hemsworth}},
  \bibinfo{author}{\bibfnamefont{A.}~\bibnamefont{Tanga}}, \bibnamefont{and}
  \bibinfo{author}{\bibfnamefont{V.}~\bibnamefont{Antoni}},
  \bibinfo{journal}{Review of Scientific Instruments}
  \textbf{\bibinfo{volume}{79}} (\bibinfo{year}{2008}).

\bibitem[{\citenamefont{Adam}(1987)}]{adam1987review}
\bibinfo{author}{\bibfnamefont{J.}~\bibnamefont{Adam}},
  \bibinfo{journal}{Plasma physics and controlled fusion}
  \textbf{\bibinfo{volume}{29}}, \bibinfo{pages}{443} (\bibinfo{year}{1987}).

\bibitem[{\citenamefont{Kazakov et~al.}(2017)\citenamefont{Kazakov, Ongena,
  Wright, Wukitch, Lerche, Mantsinen, Van~Eester, Craciunescu, Kiptily, Lin
  et~al.}}]{kazakov2017efficient}
\bibinfo{author}{\bibfnamefont{Y.~O.} \bibnamefont{Kazakov}},
  \bibinfo{author}{\bibfnamefont{J.}~\bibnamefont{Ongena}},
  \bibinfo{author}{\bibfnamefont{J.}~\bibnamefont{Wright}},
  \bibinfo{author}{\bibfnamefont{S.}~\bibnamefont{Wukitch}},
  \bibinfo{author}{\bibfnamefont{E.}~\bibnamefont{Lerche}},
  \bibinfo{author}{\bibfnamefont{M.}~\bibnamefont{Mantsinen}},
  \bibinfo{author}{\bibfnamefont{D.}~\bibnamefont{Van~Eester}},
  \bibinfo{author}{\bibfnamefont{T.}~\bibnamefont{Craciunescu}},
  \bibinfo{author}{\bibfnamefont{V.}~\bibnamefont{Kiptily}},
  \bibinfo{author}{\bibfnamefont{Y.}~\bibnamefont{Lin}}, \bibnamefont{et~al.},
  \bibinfo{journal}{Nature Physics} \textbf{\bibinfo{volume}{13}},
  \bibinfo{pages}{973} (\bibinfo{year}{2017}).

\bibitem[{\citenamefont{Fasoli et~al.}(2007)\citenamefont{Fasoli, Gormenzano,
  Berk, Breizman, Briguglio, Darrow, Gorelenkov, Heidbrink, Jaun, Konovalov
  et~al.}}]{fasoli2007physics}
\bibinfo{author}{\bibfnamefont{A.}~\bibnamefont{Fasoli}},
  \bibinfo{author}{\bibfnamefont{C.}~\bibnamefont{Gormenzano}},
  \bibinfo{author}{\bibfnamefont{H.}~\bibnamefont{Berk}},
  \bibinfo{author}{\bibfnamefont{B.}~\bibnamefont{Breizman}},
  \bibinfo{author}{\bibfnamefont{S.}~\bibnamefont{Briguglio}},
  \bibinfo{author}{\bibfnamefont{D.}~\bibnamefont{Darrow}},
  \bibinfo{author}{\bibfnamefont{N.}~\bibnamefont{Gorelenkov}},
  \bibinfo{author}{\bibfnamefont{W.}~\bibnamefont{Heidbrink}},
  \bibinfo{author}{\bibfnamefont{A.}~\bibnamefont{Jaun}},
  \bibinfo{author}{\bibfnamefont{S.}~\bibnamefont{Konovalov}},
  \bibnamefont{et~al.}, \bibinfo{journal}{Nuclear Fusion}
  \textbf{\bibinfo{volume}{47}}, \bibinfo{pages}{S264} (\bibinfo{year}{2007}).

\bibitem[{\citenamefont{Bonofiglo et~al.}(2024)\citenamefont{Bonofiglo,
  Kiptily, Rivero-Rodriguez, Nocente, Podesta, {\v{S}}tancar, Poradzinski,
  Goloborodko, Sharapov, Fitzgerald et~al.}}]{bonofiglo2024alpha}
\bibinfo{author}{\bibfnamefont{P.}~\bibnamefont{Bonofiglo}},
  \bibinfo{author}{\bibfnamefont{V.}~\bibnamefont{Kiptily}},
  \bibinfo{author}{\bibfnamefont{J.}~\bibnamefont{Rivero-Rodriguez}},
  \bibinfo{author}{\bibfnamefont{M.}~\bibnamefont{Nocente}},
  \bibinfo{author}{\bibfnamefont{M.}~\bibnamefont{Podesta}},
  \bibinfo{author}{\bibfnamefont{{\v{Z}}.}~\bibnamefont{{\v{S}}tancar}},
  \bibinfo{author}{\bibfnamefont{M.}~\bibnamefont{Poradzinski}},
  \bibinfo{author}{\bibfnamefont{V.}~\bibnamefont{Goloborodko}},
  \bibinfo{author}{\bibfnamefont{S.~E.} \bibnamefont{Sharapov}},
  \bibinfo{author}{\bibfnamefont{M.}~\bibnamefont{Fitzgerald}},
  \bibnamefont{et~al.}, \bibinfo{journal}{Nuclear Fusion}
  \textbf{\bibinfo{volume}{64}}, \bibinfo{pages}{096038}
  (\bibinfo{year}{2024}).

\bibitem[{\citenamefont{Goldston et~al.}(1981)\citenamefont{Goldston, White,
  and Boozer}}]{goldston1981confinement}
\bibinfo{author}{\bibfnamefont{R.~J.} \bibnamefont{Goldston}},
  \bibinfo{author}{\bibfnamefont{R.}~\bibnamefont{White}}, \bibnamefont{and}
  \bibinfo{author}{\bibfnamefont{A.~H.} \bibnamefont{Boozer}},
  \bibinfo{journal}{Physical review letters} \textbf{\bibinfo{volume}{47}},
  \bibinfo{pages}{647} (\bibinfo{year}{1981}).

\bibitem[{\citenamefont{Carolipio et~al.}(2002)\citenamefont{Carolipio,
  Heidbrink, Forest, and White}}]{carolipio2002simulations}
\bibinfo{author}{\bibfnamefont{E.}~\bibnamefont{Carolipio}},
  \bibinfo{author}{\bibfnamefont{W.}~\bibnamefont{Heidbrink}},
  \bibinfo{author}{\bibfnamefont{C.}~\bibnamefont{Forest}}, \bibnamefont{and}
  \bibinfo{author}{\bibfnamefont{R.}~\bibnamefont{White}},
  \bibinfo{journal}{Nuclear fusion} \textbf{\bibinfo{volume}{42}},
  \bibinfo{pages}{853} (\bibinfo{year}{2002}).

\bibitem[{\citenamefont{Heidbrink and White}(2020)}]{heidbrink2020mechanisms}
\bibinfo{author}{\bibfnamefont{W.}~\bibnamefont{Heidbrink}} \bibnamefont{and}
  \bibinfo{author}{\bibfnamefont{R.}~\bibnamefont{White}},
  \bibinfo{journal}{Physics of Plasmas} \textbf{\bibinfo{volume}{27}}
  (\bibinfo{year}{2020}).

\bibitem[{\citenamefont{Galeev et~al.}(1969)\citenamefont{Galeev, Sagdeev,
  Furth, and Rosenbluth}}]{galeev1969plasma}
\bibinfo{author}{\bibfnamefont{A.~A.} \bibnamefont{Galeev}},
  \bibinfo{author}{\bibfnamefont{R.}~\bibnamefont{Sagdeev}},
  \bibinfo{author}{\bibfnamefont{H.}~\bibnamefont{Furth}}, \bibnamefont{and}
  \bibinfo{author}{\bibfnamefont{M.}~\bibnamefont{Rosenbluth}},
  \bibinfo{journal}{Physical Review Letters} \textbf{\bibinfo{volume}{22}},
  \bibinfo{pages}{511} (\bibinfo{year}{1969}).

\bibitem[{\citenamefont{S{\"a}rkim{\"a}ki}(2020)}]{sarkimaki2020efficient}
\bibinfo{author}{\bibfnamefont{K.}~\bibnamefont{S{\"a}rkim{\"a}ki}},
  \bibinfo{journal}{Nuclear Fusion} \textbf{\bibinfo{volume}{60}},
  \bibinfo{pages}{036002} (\bibinfo{year}{2020}).

\bibitem[{\citenamefont{Raissi et~al.}(2019)\citenamefont{Raissi, Perdikaris,
  and Karniadakis}}]{raissi2019physics}
\bibinfo{author}{\bibfnamefont{M.}~\bibnamefont{Raissi}},
  \bibinfo{author}{\bibfnamefont{P.}~\bibnamefont{Perdikaris}},
  \bibnamefont{and} \bibinfo{author}{\bibfnamefont{G.~E.}
  \bibnamefont{Karniadakis}}, \bibinfo{journal}{Journal of Computational
  physics} \textbf{\bibinfo{volume}{378}}, \bibinfo{pages}{686}
  (\bibinfo{year}{2019}).

\bibitem[{\citenamefont{Sun et~al.}(2020)\citenamefont{Sun, Gao, Pan, and
  Wang}}]{sun2020surrogate}
\bibinfo{author}{\bibfnamefont{L.}~\bibnamefont{Sun}},
  \bibinfo{author}{\bibfnamefont{H.}~\bibnamefont{Gao}},
  \bibinfo{author}{\bibfnamefont{S.}~\bibnamefont{Pan}}, \bibnamefont{and}
  \bibinfo{author}{\bibfnamefont{J.-X.} \bibnamefont{Wang}},
  \bibinfo{journal}{Computer Methods in Applied Mechanics and Engineering}
  \textbf{\bibinfo{volume}{361}}, \bibinfo{pages}{112732}
  (\bibinfo{year}{2020}).

\bibitem[{\citenamefont{Arnaud et~al.}(2024)\citenamefont{Arnaud, Mark, and
  McDevitt}}]{arnaud2024physics}
\bibinfo{author}{\bibfnamefont{J.~S.} \bibnamefont{Arnaud}},
  \bibinfo{author}{\bibfnamefont{T.}~\bibnamefont{Mark}}, \bibnamefont{and}
  \bibinfo{author}{\bibfnamefont{C.~J.} \bibnamefont{McDevitt}},
  \bibinfo{journal}{J. Plasma Phys.} \textbf{\bibinfo{volume}{90}},
  \bibinfo{pages}{905900409} (\bibinfo{year}{2024}).

\bibitem[{\citenamefont{McDevitt et~al.}(2025)\citenamefont{McDevitt, Arnaud,
  and Tang}}]{mcdevitt2025primary}
\bibinfo{author}{\bibfnamefont{C.~J.} \bibnamefont{McDevitt}},
  \bibinfo{author}{\bibfnamefont{J.~S.} \bibnamefont{Arnaud}},
  \bibnamefont{and} \bibinfo{author}{\bibfnamefont{X.-Z.} \bibnamefont{Tang}},
  \bibinfo{journal}{Physics of Plasmas} \textbf{\bibinfo{volume}{32}}
  (\bibinfo{year}{2025}).

\bibitem[{\citenamefont{Arnaud et~al.}(2025)\citenamefont{Arnaud, Tang, and
  McDevitt}}]{Arnaud2025arunaway}
\bibinfo{author}{\bibfnamefont{J.~S.} \bibnamefont{Arnaud}},
  \bibinfo{author}{\bibfnamefont{X.-Z.} \bibnamefont{Tang}}, \bibnamefont{and}
  \bibinfo{author}{\bibfnamefont{C.~J.} \bibnamefont{McDevitt}},
  \bibinfo{journal}{Nuclear Fusion} \textbf{\bibinfo{volume}{65}},
  \bibinfo{pages}{106013} (\bibinfo{year}{2025}).

\bibitem[{\citenamefont{McDevitt and Tang}(2024)}]{mcdevitt2024knudsen}
\bibinfo{author}{\bibfnamefont{C.~J.} \bibnamefont{McDevitt}} \bibnamefont{and}
  \bibinfo{author}{\bibfnamefont{X.-Z.} \bibnamefont{Tang}},
  \bibinfo{journal}{Physics of Plasmas} \textbf{\bibinfo{volume}{31}}
  (\bibinfo{year}{2024}).

\bibitem[{\citenamefont{Urb{\'a}n et~al.}(2025)\citenamefont{Urb{\'a}n,
  Stefanou, and Pons}}]{urban2025unveiling}
\bibinfo{author}{\bibfnamefont{J.~F.} \bibnamefont{Urb{\'a}n}},
  \bibinfo{author}{\bibfnamefont{P.}~\bibnamefont{Stefanou}}, \bibnamefont{and}
  \bibinfo{author}{\bibfnamefont{J.~A.} \bibnamefont{Pons}},
  \bibinfo{journal}{Journal of Computational Physics}
  \textbf{\bibinfo{volume}{523}}, \bibinfo{pages}{113656}
  (\bibinfo{year}{2025}).

\bibitem[{\citenamefont{Kiyani et~al.}(2025)\citenamefont{Kiyani, Shukla,
  Urb{\'a}n, Darbon, and Karniadakis}}]{kiyani2025optimizer}
\bibinfo{author}{\bibfnamefont{E.}~\bibnamefont{Kiyani}},
  \bibinfo{author}{\bibfnamefont{K.}~\bibnamefont{Shukla}},
  \bibinfo{author}{\bibfnamefont{J.~F.} \bibnamefont{Urb{\'a}n}},
  \bibinfo{author}{\bibfnamefont{J.}~\bibnamefont{Darbon}}, \bibnamefont{and}
  \bibinfo{author}{\bibfnamefont{G.~E.} \bibnamefont{Karniadakis}},
  \bibinfo{journal}{Computer Methods in Applied Mechanics and Engineering}
  \textbf{\bibinfo{volume}{446}}, \bibinfo{pages}{118308}
  (\bibinfo{year}{2025}).

\bibitem[{\citenamefont{Wang et~al.}(2025)\citenamefont{Wang, Bhartari, Li, and
  Perdikaris}}]{wang2025gradient}
\bibinfo{author}{\bibfnamefont{S.}~\bibnamefont{Wang}},
  \bibinfo{author}{\bibfnamefont{A.~K.} \bibnamefont{Bhartari}},
  \bibinfo{author}{\bibfnamefont{B.}~\bibnamefont{Li}}, \bibnamefont{and}
  \bibinfo{author}{\bibfnamefont{P.}~\bibnamefont{Perdikaris}},
  \bibinfo{journal}{arXiv preprint arXiv:2502.00604}  (\bibinfo{year}{2025}).

\bibitem[{\citenamefont{Hinton and Chu}(1985)}]{hinton1985neoclassical}
\bibinfo{author}{\bibfnamefont{F.}~\bibnamefont{Hinton}} \bibnamefont{and}
  \bibinfo{author}{\bibfnamefont{M.}~\bibnamefont{Chu}},
  \bibinfo{journal}{Nuclear fusion} \textbf{\bibinfo{volume}{25}},
  \bibinfo{pages}{345} (\bibinfo{year}{1985}).

\bibitem[{\citenamefont{Helander and Sigmar}(2002)}]{Helander-Sigmar:book}
\bibinfo{author}{\bibfnamefont{P.}~\bibnamefont{Helander}} \bibnamefont{and}
  \bibinfo{author}{\bibfnamefont{D.~J.} \bibnamefont{Sigmar}},
  \emph{\bibinfo{title}{Collisional Transport in Magnetized Plasmas}}
  (\bibinfo{publisher}{Cambridge University Press, Cambridge},
  \bibinfo{year}{2002}).

\bibitem[{\citenamefont{Grasman et~al.}(2013)\citenamefont{Grasman, Onno
  et~al.}}]{grasman2013asymptotic}
\bibinfo{author}{\bibfnamefont{J.}~\bibnamefont{Grasman}},
  \bibinfo{author}{\bibfnamefont{A.}~\bibnamefont{Onno}}, \bibnamefont{et~al.},
  \emph{\bibinfo{title}{Asymptotic methods for the Fokker-Planck equation and
  the exit problem in applications}} (\bibinfo{publisher}{Springer Science \&
  Business Media}, \bibinfo{year}{2013}).

\bibitem[{\citenamefont{Liu et~al.}(2017)\citenamefont{Liu, Brennan, Boozer,
  and Bhattacharjee}}]{Liu:2017}
\bibinfo{author}{\bibfnamefont{C.}~\bibnamefont{Liu}},
  \bibinfo{author}{\bibfnamefont{D.~P.} \bibnamefont{Brennan}},
  \bibinfo{author}{\bibfnamefont{A.~H.} \bibnamefont{Boozer}},
  \bibnamefont{and}
  \bibinfo{author}{\bibfnamefont{A.}~\bibnamefont{Bhattacharjee}},
  \bibinfo{journal}{Plasma Physics and Controlled Fusion}
  \textbf{\bibinfo{volume}{59}}, \bibinfo{pages}{024003}
  (\bibinfo{year}{2017}).

\bibitem[{\citenamefont{Cuomo et~al.}(2022)\citenamefont{Cuomo, Di~Cola,
  Giampaolo, Rozza, Raissi, and Piccialli}}]{cuomo2022scientific}
\bibinfo{author}{\bibfnamefont{S.}~\bibnamefont{Cuomo}},
  \bibinfo{author}{\bibfnamefont{V.~S.} \bibnamefont{Di~Cola}},
  \bibinfo{author}{\bibfnamefont{F.}~\bibnamefont{Giampaolo}},
  \bibinfo{author}{\bibfnamefont{G.}~\bibnamefont{Rozza}},
  \bibinfo{author}{\bibfnamefont{M.}~\bibnamefont{Raissi}}, \bibnamefont{and}
  \bibinfo{author}{\bibfnamefont{F.}~\bibnamefont{Piccialli}},
  \bibinfo{journal}{Journal of Scientific Computing}
  \textbf{\bibinfo{volume}{92}}, \bibinfo{pages}{88} (\bibinfo{year}{2022}).

\bibitem[{\citenamefont{Wang et~al.}(2023)\citenamefont{Wang, Sankaran, Wang,
  and Perdikaris}}]{wang2023expert}
\bibinfo{author}{\bibfnamefont{S.}~\bibnamefont{Wang}},
  \bibinfo{author}{\bibfnamefont{S.}~\bibnamefont{Sankaran}},
  \bibinfo{author}{\bibfnamefont{H.}~\bibnamefont{Wang}}, \bibnamefont{and}
  \bibinfo{author}{\bibfnamefont{P.}~\bibnamefont{Perdikaris}},
  \bibinfo{journal}{arXiv preprint arXiv:2308.08468}  (\bibinfo{year}{2023}).

\bibitem[{\citenamefont{Toscano et~al.}(2025)\citenamefont{Toscano, Oommen,
  Varghese, Zou, Ahmadi~Daryakenari, Wu, and Karniadakis}}]{toscano2025pinns}
\bibinfo{author}{\bibfnamefont{J.~D.} \bibnamefont{Toscano}},
  \bibinfo{author}{\bibfnamefont{V.}~\bibnamefont{Oommen}},
  \bibinfo{author}{\bibfnamefont{A.~J.} \bibnamefont{Varghese}},
  \bibinfo{author}{\bibfnamefont{Z.}~\bibnamefont{Zou}},
  \bibinfo{author}{\bibfnamefont{N.}~\bibnamefont{Ahmadi~Daryakenari}},
  \bibinfo{author}{\bibfnamefont{C.}~\bibnamefont{Wu}}, \bibnamefont{and}
  \bibinfo{author}{\bibfnamefont{G.~E.} \bibnamefont{Karniadakis}},
  \bibinfo{journal}{Machine Learning for Computational Science and Engineering}
  \textbf{\bibinfo{volume}{1}}, \bibinfo{pages}{1} (\bibinfo{year}{2025}).

\bibitem[{\citenamefont{Karniadakis et~al.}(2021)\citenamefont{Karniadakis,
  Kevrekidis, Lu, Perdikaris, Wang, and Yang}}]{karniadakis2021physics}
\bibinfo{author}{\bibfnamefont{G.~E.} \bibnamefont{Karniadakis}},
  \bibinfo{author}{\bibfnamefont{I.~G.} \bibnamefont{Kevrekidis}},
  \bibinfo{author}{\bibfnamefont{L.}~\bibnamefont{Lu}},
  \bibinfo{author}{\bibfnamefont{P.}~\bibnamefont{Perdikaris}},
  \bibinfo{author}{\bibfnamefont{S.}~\bibnamefont{Wang}}, \bibnamefont{and}
  \bibinfo{author}{\bibfnamefont{L.}~\bibnamefont{Yang}},
  \bibinfo{journal}{Nature Reviews Physics} \textbf{\bibinfo{volume}{3}},
  \bibinfo{pages}{422} (\bibinfo{year}{2021}).

\bibitem[{\citenamefont{van Milligen et~al.}(1995)\citenamefont{van Milligen,
  Tribaldos, and Jim{\'e}nez}}]{van1995neural}
\bibinfo{author}{\bibfnamefont{B.~P.} \bibnamefont{van Milligen}},
  \bibinfo{author}{\bibfnamefont{V.}~\bibnamefont{Tribaldos}},
  \bibnamefont{and}
  \bibinfo{author}{\bibfnamefont{J.}~\bibnamefont{Jim{\'e}nez}},
  \bibinfo{journal}{Physical review letters} \textbf{\bibinfo{volume}{75}},
  \bibinfo{pages}{3594} (\bibinfo{year}{1995}).

\bibitem[{\citenamefont{Cai et~al.}(2021)\citenamefont{Cai, Mao, Wang, Yin, and
  Karniadakis}}]{cai2021physics}
\bibinfo{author}{\bibfnamefont{S.}~\bibnamefont{Cai}},
  \bibinfo{author}{\bibfnamefont{Z.}~\bibnamefont{Mao}},
  \bibinfo{author}{\bibfnamefont{Z.}~\bibnamefont{Wang}},
  \bibinfo{author}{\bibfnamefont{M.}~\bibnamefont{Yin}}, \bibnamefont{and}
  \bibinfo{author}{\bibfnamefont{G.~E.} \bibnamefont{Karniadakis}},
  \bibinfo{journal}{Acta Mechanica Sinica} \textbf{\bibinfo{volume}{37}},
  \bibinfo{pages}{1727} (\bibinfo{year}{2021}).

\bibitem[{\citenamefont{Mathews et~al.}(2022)\citenamefont{Mathews, Hughes,
  Terry, and Baek}}]{mathews2022deep}
\bibinfo{author}{\bibfnamefont{A.}~\bibnamefont{Mathews}},
  \bibinfo{author}{\bibfnamefont{J.~W.} \bibnamefont{Hughes}},
  \bibinfo{author}{\bibfnamefont{J.~L.} \bibnamefont{Terry}}, \bibnamefont{and}
  \bibinfo{author}{\bibfnamefont{S.-G.} \bibnamefont{Baek}},
  \bibinfo{journal}{Physical Review Letters} \textbf{\bibinfo{volume}{129}},
  \bibinfo{pages}{235002} (\bibinfo{year}{2022}).

\bibitem[{\citenamefont{McDevitt et~al.}(2024)\citenamefont{McDevitt, Fowler,
  and Roy}}]{mcdevittSciTech2024}
\bibinfo{author}{\bibfnamefont{C.}~\bibnamefont{McDevitt}},
  \bibinfo{author}{\bibfnamefont{E.}~\bibnamefont{Fowler}}, \bibnamefont{and}
  \bibinfo{author}{\bibfnamefont{S.}~\bibnamefont{Roy}}, in
  \emph{\bibinfo{booktitle}{AIAA SCITECH 2024 Forum}} (\bibinfo{year}{2024}),
  p. \bibinfo{pages}{1692}.

\bibitem[{\citenamefont{McDevitt}(2023)}]{McDevitt:hottail:2023}
\bibinfo{author}{\bibfnamefont{C.~J.} \bibnamefont{McDevitt}},
  \bibinfo{journal}{Physics of Plasmas} \textbf{\bibinfo{volume}{30}},
  \bibinfo{pages}{092501} (\bibinfo{year}{2023}).

\bibitem[{\citenamefont{Jang et~al.}(2024)\citenamefont{Jang, Kaptanoglu, Gaur,
  Pan, Landreman, and Dorland}}]{jang2024grad}
\bibinfo{author}{\bibfnamefont{B.}~\bibnamefont{Jang}},
  \bibinfo{author}{\bibfnamefont{A.~A.} \bibnamefont{Kaptanoglu}},
  \bibinfo{author}{\bibfnamefont{R.}~\bibnamefont{Gaur}},
  \bibinfo{author}{\bibfnamefont{S.}~\bibnamefont{Pan}},
  \bibinfo{author}{\bibfnamefont{M.}~\bibnamefont{Landreman}},
  \bibnamefont{and} \bibinfo{author}{\bibfnamefont{W.}~\bibnamefont{Dorland}},
  \bibinfo{journal}{Physics of Plasmas} \textbf{\bibinfo{volume}{31}}
  (\bibinfo{year}{2024}).

\bibitem[{\citenamefont{Luo et~al.}(2025)\citenamefont{Luo, Ren, Chen, Mao,
  Zhang, Wang, and Tang}}]{luo2025parametric}
\bibinfo{author}{\bibfnamefont{W.}~\bibnamefont{Luo}},
  \bibinfo{author}{\bibfnamefont{J.-x.} \bibnamefont{Ren}},
  \bibinfo{author}{\bibfnamefont{Z.}~\bibnamefont{Chen}},
  \bibinfo{author}{\bibfnamefont{R.}~\bibnamefont{Mao}},
  \bibinfo{author}{\bibfnamefont{G.}~\bibnamefont{Zhang}},
  \bibinfo{author}{\bibfnamefont{Y.}~\bibnamefont{Wang}}, \bibnamefont{and}
  \bibinfo{author}{\bibfnamefont{H.}~\bibnamefont{Tang}},
  \bibinfo{journal}{Physics of Fluids} \textbf{\bibinfo{volume}{37}}
  (\bibinfo{year}{2025}).

\bibitem[{\citenamefont{Kingma and Ba}(2014)}]{kingma2014adam}
\bibinfo{author}{\bibfnamefont{D.~P.} \bibnamefont{Kingma}} \bibnamefont{and}
  \bibinfo{author}{\bibfnamefont{J.}~\bibnamefont{Ba}}, \bibinfo{journal}{arXiv
  preprint arXiv:1412.6980}  (\bibinfo{year}{2014}).

\bibitem[{\citenamefont{Liu and Nocedal}(1989)}]{liu1989limited}
\bibinfo{author}{\bibfnamefont{D.~C.} \bibnamefont{Liu}} \bibnamefont{and}
  \bibinfo{author}{\bibfnamefont{J.}~\bibnamefont{Nocedal}},
  \bibinfo{journal}{Mathematical programming} \textbf{\bibinfo{volume}{45}},
  \bibinfo{pages}{503} (\bibinfo{year}{1989}).

\bibitem[{\citenamefont{Vyas et~al.}(2024)\citenamefont{Vyas, Morwani, Zhao,
  Kwun, Shapira, Brandfonbrener, Janson, and Kakade}}]{vyas2024soap}
\bibinfo{author}{\bibfnamefont{N.}~\bibnamefont{Vyas}},
  \bibinfo{author}{\bibfnamefont{D.}~\bibnamefont{Morwani}},
  \bibinfo{author}{\bibfnamefont{R.}~\bibnamefont{Zhao}},
  \bibinfo{author}{\bibfnamefont{M.}~\bibnamefont{Kwun}},
  \bibinfo{author}{\bibfnamefont{I.}~\bibnamefont{Shapira}},
  \bibinfo{author}{\bibfnamefont{D.}~\bibnamefont{Brandfonbrener}},
  \bibinfo{author}{\bibfnamefont{L.}~\bibnamefont{Janson}}, \bibnamefont{and}
  \bibinfo{author}{\bibfnamefont{S.}~\bibnamefont{Kakade}},
  \bibinfo{journal}{arXiv preprint arXiv:2409.11321}  (\bibinfo{year}{2024}).

\bibitem[{\citenamefont{Al-Baali et~al.}(2014)\citenamefont{Al-Baali,
  Spedicato, and Maggioni}}]{al2014broyden}
\bibinfo{author}{\bibfnamefont{M.}~\bibnamefont{Al-Baali}},
  \bibinfo{author}{\bibfnamefont{E.}~\bibnamefont{Spedicato}},
  \bibnamefont{and} \bibinfo{author}{\bibfnamefont{F.}~\bibnamefont{Maggioni}},
  \bibinfo{journal}{Optimization Methods and Software}
  \textbf{\bibinfo{volume}{29}}, \bibinfo{pages}{937} (\bibinfo{year}{2014}).

\bibitem[{\citenamefont{White and Chance}(1984)}]{white1984hamiltonian}
\bibinfo{author}{\bibfnamefont{R.~t.} \bibnamefont{White}} \bibnamefont{and}
  \bibinfo{author}{\bibfnamefont{M.}~\bibnamefont{Chance}},
  \bibinfo{type}{Tech. Rep.}, \bibinfo{institution}{Princeton Plasma Physics
  Lab.(PPPL), Princeton, NJ (United States)} (\bibinfo{year}{1984}).

\bibitem[{\citenamefont{Hirvijoki et~al.}(2014)\citenamefont{Hirvijoki, Asunta,
  Koskela, Kurki-Suonio, Miettunen, Sipil{\"a}, Snicker, and
  {\"A}k{\"a}slompolo}}]{hirvijoki2014ascot}
\bibinfo{author}{\bibfnamefont{E.}~\bibnamefont{Hirvijoki}},
  \bibinfo{author}{\bibfnamefont{O.}~\bibnamefont{Asunta}},
  \bibinfo{author}{\bibfnamefont{T.}~\bibnamefont{Koskela}},
  \bibinfo{author}{\bibfnamefont{T.}~\bibnamefont{Kurki-Suonio}},
  \bibinfo{author}{\bibfnamefont{J.}~\bibnamefont{Miettunen}},
  \bibinfo{author}{\bibfnamefont{S.}~\bibnamefont{Sipil{\"a}}},
  \bibinfo{author}{\bibfnamefont{A.}~\bibnamefont{Snicker}}, \bibnamefont{and}
  \bibinfo{author}{\bibfnamefont{S.}~\bibnamefont{{\"A}k{\"a}slompolo}},
  \bibinfo{journal}{Computer Physics Communications}
  \textbf{\bibinfo{volume}{185}}, \bibinfo{pages}{1310} (\bibinfo{year}{2014}).

\bibitem[{\citenamefont{Trott et~al.}(2021)\citenamefont{Trott,
  Lebrun-Grandi{\'e}, Arndt, Ciesko, Dang, Ellingwood, Gayatri, Harvey,
  Hollman, Ibanez et~al.}}]{trott2021kokkos}
\bibinfo{author}{\bibfnamefont{C.~R.} \bibnamefont{Trott}},
  \bibinfo{author}{\bibfnamefont{D.}~\bibnamefont{Lebrun-Grandi{\'e}}},
  \bibinfo{author}{\bibfnamefont{D.}~\bibnamefont{Arndt}},
  \bibinfo{author}{\bibfnamefont{J.}~\bibnamefont{Ciesko}},
  \bibinfo{author}{\bibfnamefont{V.}~\bibnamefont{Dang}},
  \bibinfo{author}{\bibfnamefont{N.}~\bibnamefont{Ellingwood}},
  \bibinfo{author}{\bibfnamefont{R.}~\bibnamefont{Gayatri}},
  \bibinfo{author}{\bibfnamefont{E.}~\bibnamefont{Harvey}},
  \bibinfo{author}{\bibfnamefont{D.~S.} \bibnamefont{Hollman}},
  \bibinfo{author}{\bibfnamefont{D.}~\bibnamefont{Ibanez}},
  \bibnamefont{et~al.}, \bibinfo{journal}{IEEE Transactions on Parallel and
  Distributed Systems} \textbf{\bibinfo{volume}{33}}, \bibinfo{pages}{805}
  (\bibinfo{year}{2021}).

\bibitem[{\citenamefont{Paszke et~al.}(2019)\citenamefont{Paszke, Gross, Massa,
  Lerer, Bradbury, Chanan, Killeen, Lin, Gimelshein, Antiga
  et~al.}}]{paszke2019pytorch}
\bibinfo{author}{\bibfnamefont{A.}~\bibnamefont{Paszke}},
  \bibinfo{author}{\bibfnamefont{S.}~\bibnamefont{Gross}},
  \bibinfo{author}{\bibfnamefont{F.}~\bibnamefont{Massa}},
  \bibinfo{author}{\bibfnamefont{A.}~\bibnamefont{Lerer}},
  \bibinfo{author}{\bibfnamefont{J.}~\bibnamefont{Bradbury}},
  \bibinfo{author}{\bibfnamefont{G.}~\bibnamefont{Chanan}},
  \bibinfo{author}{\bibfnamefont{T.}~\bibnamefont{Killeen}},
  \bibinfo{author}{\bibfnamefont{Z.}~\bibnamefont{Lin}},
  \bibinfo{author}{\bibfnamefont{N.}~\bibnamefont{Gimelshein}},
  \bibinfo{author}{\bibfnamefont{L.}~\bibnamefont{Antiga}},
  \bibnamefont{et~al.}, \bibinfo{journal}{Advances in neural information
  processing systems} \textbf{\bibinfo{volume}{32}} (\bibinfo{year}{2019}).

\bibitem[{\citenamefont{Bradbury et~al.}(2018)\citenamefont{Bradbury, Frostig,
  Hawkins, Johnson, Leary, Maclaurin, Necula, Paszke, VanderPlas,
  Wanderman-Milne et~al.}}]{bradbury2018jax}
\bibinfo{author}{\bibfnamefont{J.}~\bibnamefont{Bradbury}},
  \bibinfo{author}{\bibfnamefont{R.}~\bibnamefont{Frostig}},
  \bibinfo{author}{\bibfnamefont{P.}~\bibnamefont{Hawkins}},
  \bibinfo{author}{\bibfnamefont{M.~J.} \bibnamefont{Johnson}},
  \bibinfo{author}{\bibfnamefont{C.}~\bibnamefont{Leary}},
  \bibinfo{author}{\bibfnamefont{D.}~\bibnamefont{Maclaurin}},
  \bibinfo{author}{\bibfnamefont{G.}~\bibnamefont{Necula}},
  \bibinfo{author}{\bibfnamefont{A.}~\bibnamefont{Paszke}},
  \bibinfo{author}{\bibfnamefont{J.}~\bibnamefont{VanderPlas}},
  \bibinfo{author}{\bibfnamefont{S.}~\bibnamefont{Wanderman-Milne}},
  \bibnamefont{et~al.} (\bibinfo{year}{2018}).

\bibitem[{opt(2020)}]{optax2020}
\emph{\bibinfo{title}{Optax: Gradient processing and optimization library in
  jax}} (\bibinfo{year}{2020}),
  \urlprefix\url{https://github.com/deepmind/optax}.

\bibitem[{\citenamefont{Boozer and Kuo-Petravic}(1981)}]{Boozer:1981}
\bibinfo{author}{\bibfnamefont{A.~H.} \bibnamefont{Boozer}} \bibnamefont{and}
  \bibinfo{author}{\bibfnamefont{G.}~\bibnamefont{Kuo-Petravic}},
  \bibinfo{journal}{The Physics of Fluids} \textbf{\bibinfo{volume}{24}},
  \bibinfo{pages}{851} (\bibinfo{year}{1981}).

\bibitem[{\citenamefont{Rosenbluth et~al.}(1972)\citenamefont{Rosenbluth,
  Hazeltine, and Hinton}}]{Rosenbluth:1972}
\bibinfo{author}{\bibfnamefont{M.}~\bibnamefont{Rosenbluth}},
  \bibinfo{author}{\bibfnamefont{R.}~\bibnamefont{Hazeltine}},
  \bibnamefont{and} \bibinfo{author}{\bibfnamefont{F.~L.}
  \bibnamefont{Hinton}}, \bibinfo{journal}{The Physics of Fluids}
  \textbf{\bibinfo{volume}{15}}, \bibinfo{pages}{116} (\bibinfo{year}{1972}).

\bibitem[{\citenamefont{Wu et~al.}(2023)\citenamefont{Wu, Zhu, Tan, Kartha, and
  Lu}}]{wu2023comprehensive}
\bibinfo{author}{\bibfnamefont{C.}~\bibnamefont{Wu}},
  \bibinfo{author}{\bibfnamefont{M.}~\bibnamefont{Zhu}},
  \bibinfo{author}{\bibfnamefont{Q.}~\bibnamefont{Tan}},
  \bibinfo{author}{\bibfnamefont{Y.}~\bibnamefont{Kartha}}, \bibnamefont{and}
  \bibinfo{author}{\bibfnamefont{L.}~\bibnamefont{Lu}},
  \bibinfo{journal}{Computer Methods in Applied Mechanics and Engineering}
  \textbf{\bibinfo{volume}{403}}, \bibinfo{pages}{115671}
  (\bibinfo{year}{2023}).

\bibitem[{\citenamefont{Law et~al.}(2022)\citenamefont{Law, Cerfon, and
  Peherstorfer}}]{law2022accelerating}
\bibinfo{author}{\bibfnamefont{F.}~\bibnamefont{Law}},
  \bibinfo{author}{\bibfnamefont{A.}~\bibnamefont{Cerfon}}, \bibnamefont{and}
  \bibinfo{author}{\bibfnamefont{B.}~\bibnamefont{Peherstorfer}},
  \bibinfo{journal}{Nuclear Fusion} \textbf{\bibinfo{volume}{62}},
  \bibinfo{pages}{076019} (\bibinfo{year}{2022}).

\bibitem[{\citenamefont{Kaltsas and
  Throumoulopoulos}(2022)}]{kaltsas2022neural}
\bibinfo{author}{\bibfnamefont{D.}~\bibnamefont{Kaltsas}} \bibnamefont{and}
  \bibinfo{author}{\bibfnamefont{G.}~\bibnamefont{Throumoulopoulos}},
  \bibinfo{journal}{Physics of Plasmas} \textbf{\bibinfo{volume}{29}}
  (\bibinfo{year}{2022}).

\bibitem[{\citenamefont{Brzozowski et~al.}(2019)\citenamefont{Brzozowski,
  Jenko, Bilato, Cavedon, Team et~al.}}]{brzozowski2019geometric}
\bibinfo{author}{\bibfnamefont{R.~W.} \bibnamefont{Brzozowski}},
  \bibinfo{author}{\bibfnamefont{F.}~\bibnamefont{Jenko}},
  \bibinfo{author}{\bibfnamefont{R.}~\bibnamefont{Bilato}},
  \bibinfo{author}{\bibfnamefont{M.}~\bibnamefont{Cavedon}},
  \bibinfo{author}{\bibfnamefont{A.~U.} \bibnamefont{Team}},
  \bibnamefont{et~al.}, \bibinfo{journal}{Physics of Plasmas}
  \textbf{\bibinfo{volume}{26}} (\bibinfo{year}{2019}).

\bibitem[{\citenamefont{Salewski et~al.}(2011)\citenamefont{Salewski, Nielsen,
  Bindslev, Furtula, Gorelenkov, Korsholm, Leipold, Meo, Michelsen, Moseev
  et~al.}}]{salewski2011velocity}
\bibinfo{author}{\bibfnamefont{M.}~\bibnamefont{Salewski}},
  \bibinfo{author}{\bibfnamefont{S.~K.} \bibnamefont{Nielsen}},
  \bibinfo{author}{\bibfnamefont{H.}~\bibnamefont{Bindslev}},
  \bibinfo{author}{\bibfnamefont{V.}~\bibnamefont{Furtula}},
  \bibinfo{author}{\bibfnamefont{N.}~\bibnamefont{Gorelenkov}},
  \bibinfo{author}{\bibfnamefont{S.~B.} \bibnamefont{Korsholm}},
  \bibinfo{author}{\bibfnamefont{F.}~\bibnamefont{Leipold}},
  \bibinfo{author}{\bibfnamefont{F.}~\bibnamefont{Meo}},
  \bibinfo{author}{\bibfnamefont{P.}~\bibnamefont{Michelsen}},
  \bibinfo{author}{\bibfnamefont{D.}~\bibnamefont{Moseev}},
  \bibnamefont{et~al.}, \bibinfo{journal}{Nuclear Fusion}
  \textbf{\bibinfo{volume}{51}}, \bibinfo{pages}{083014}
  (\bibinfo{year}{2011}).

\bibitem[{\citenamefont{Rosenbluth and Putvinski}(1997)}]{Rosenbluth:1997}
\bibinfo{author}{\bibfnamefont{M.}~\bibnamefont{Rosenbluth}} \bibnamefont{and}
  \bibinfo{author}{\bibfnamefont{S.}~\bibnamefont{Putvinski}},
  \bibinfo{journal}{Nuclear Fusion} \textbf{\bibinfo{volume}{37}},
  \bibinfo{pages}{1355} (\bibinfo{year}{1997}).

\bibitem[{\citenamefont{McDevitt and Tang}(2019)}]{mcdevitt2019runaway}
\bibinfo{author}{\bibfnamefont{C.}~\bibnamefont{McDevitt}} \bibnamefont{and}
  \bibinfo{author}{\bibfnamefont{X.-Z.} \bibnamefont{Tang}},
  \bibinfo{journal}{EPL (Europhysics Letters)} \textbf{\bibinfo{volume}{127}},
  \bibinfo{pages}{45001} (\bibinfo{year}{2019}).

\bibitem[{\citenamefont{Arnaud and McDevitt}(2024)}]{arnaud2024impact}
\bibinfo{author}{\bibfnamefont{J.}~\bibnamefont{Arnaud}} \bibnamefont{and}
  \bibinfo{author}{\bibfnamefont{C.}~\bibnamefont{McDevitt}},
  \bibinfo{journal}{Physics of Plasmas} \textbf{\bibinfo{volume}{31}}
  (\bibinfo{year}{2024}).

\bibitem[{\citenamefont{Nilsson et~al.}(2015)\citenamefont{Nilsson, Decker,
  Fisch, and Peysson}}]{nilsson2015trapped}
\bibinfo{author}{\bibfnamefont{E.}~\bibnamefont{Nilsson}},
  \bibinfo{author}{\bibfnamefont{J.}~\bibnamefont{Decker}},
  \bibinfo{author}{\bibfnamefont{N.~J.} \bibnamefont{Fisch}}, \bibnamefont{and}
  \bibinfo{author}{\bibfnamefont{Y.}~\bibnamefont{Peysson}},
  \bibinfo{journal}{Journal of Plasma Physics} \textbf{\bibinfo{volume}{81}}
  (\bibinfo{year}{2015}).

\bibitem[{\citenamefont{McDevitt et~al.}(2019)\citenamefont{McDevitt, Guo, and
  Tang}}]{mcdevitt2018spatial}
\bibinfo{author}{\bibfnamefont{C.~J.} \bibnamefont{McDevitt}},
  \bibinfo{author}{\bibfnamefont{Z.}~\bibnamefont{Guo}}, \bibnamefont{and}
  \bibinfo{author}{\bibfnamefont{X.}~\bibnamefont{Tang}},
  \bibinfo{journal}{Plasma Physics and Controlled Fusion}
  \textbf{\bibinfo{volume}{61}}, \bibinfo{pages}{024004}
  (\bibinfo{year}{2019}).

\end{thebibliography}

\end{document}